\documentclass[preprint,showpacs,preprintnumbers,amsmath,amssymb,floatfix]{revtex4}

\usepackage{graphicx}
\usepackage{supertabular}
\usepackage{epsfig}
\usepackage{dcolumn}
\usepackage{bm}
\usepackage{wrapfig} 
\begin{document} 

\title{Microscopic-Macroscopic Approach for Binding Energies with 
Wigner-Kirkwood Method}
\author{A. Bhagwat$^{1}$\footnote{Electronic address: ameeya@kth.se. Present 
address: Dept. of Physics, IIT Gandhinagar, India.}, X. Vi\~nas$^{2}$,
M. Centelles$^{2}$, P. Schuck$^{3,4}$ and R. Wyss$^{1}$}
\affiliation{
$^{1}$KTH (Royal Institute of Technology), Alba Nova University Center, 
Department of Nuclear Physics, S-10691 Stockholm, Sweden\\
$^{2}$Departament d'Estructura i Constituents de la Mat\`eria
and Institut de Ci\`encies del Cosmos, Facultat de F\'{\i}sica,
Universitat de Barcelona, Diagonal {\sl 647}, {\sl E-08028} Barcelona,
Spain \\
$^{3}$Institut de Physique Nucl\'eaire, IN2P3-CNRS, Universit\'e Paris-Sud, 
F-91406 Orsay-C\'edex, France \\ 
$^{4}$ Laboratoire de Physique et Mod\'elisation des Milieux Condens\'es,
CNRS and Universit\'e Joseph Fourier, 25 Avenue des Martyrs, Bo\^{i}te Postale 166, 
F-38042 Grenoble Cedex 9, France}
\date{\today}

\begin{abstract}
The semi-classical Wigner-Kirkwood $\hbar$ expansion method is used to calculate
shell corrections for spherical and deformed nuclei. The expansion is
carried out up to fourth order in $\hbar$. A systematic study of Wigner-Kirkwood
averaged energies is presented as a function of the deformation degrees of
freedom. The shell corrections, along with the pairing energies obtained by
using the Lipkin-Nogami scheme, are used in the microscopic-macroscopic approach
to calculate binding energies. The macroscopic part is obtained from a liquid
drop formula with six adjustable parameters. Considering a set of 367
spherical nuclei, the liquid drop parameters are adjusted to reproduce the
experimental binding energies, which yields a {\it rms} deviation of 630 keV. It
is shown that the proposed approach is indeed promising for the prediction
of nuclear masses.
\end{abstract}


\maketitle

\section{Introduction}

Production and study of loosely bound exotic nuclei using Radioactive Ion Beam
facilities is of current interest \cite{limits,ENAM}. These experiments have
given rise to a number of interesting and important discoveries in nuclear
physics, like neutron and proton halos, thick skins, disappearance of magicity
at the conventional numbers and appearance of new magic numbers, etc. Further,
advances in detector systems, and in particular, the development of radioactive
beam facilities like Spiral, REX-Isolde, FAIR, and the future FRIB may allow to
investigate new features of atomic nuclei in a novel manner.

The study of nuclear masses and the systematics thereof is of immense
importance in nuclear physics. With the advent of mass spectrometry, it is
possible to measure masses of some of the short lived nuclei
spanning almost the entire periodic table \cite{LUN.03,BLA.06}. For example, 
the ISOL (isotope separator online) based mass analyzer for superheavy atoms
(MASHA) \cite{OGA.02,OGA.03} coming up at JINR-Dubna will be able to 
directly measure the masses of separated atoms in the range 
112 $\leq$ Z $\leq$ 120. The limitation on measurements is set by the 
shortest measurable half-life, $T_{1/2}~ \sim$ 1.0 s \cite{OGA.02}.
The JYFLTRAP \cite{KOL.04} developed at the University of Jyv\"askyl\"a, on
the other hand, enables to measure masses of stable as well as highly neutron 
deficient nuclei (for masses up to $A = 120$) with very high precision
($\sim$50 keV) \cite{KOL.04}. 

On the theoretical front as well, considerable progress has already been 
achieved in the accurate prediction of the nuclear masses, and it is  
still being pursued vigorously by a number of groups around the globe. 
This is of great importance, since an accurate knowledge of the nuclear 
masses plays a decisive role in a reliable description of processes like
the astrophysical r-process (see, for example, \cite{LUN.03}).
There are primarily two distinct approaches to calculate masses: 
a) the microscopic nuclear models based on density functional theory 
like, Skyrme \cite{GOR.02,GOR.09} and Gogny \cite{HIL.07} 
Hartree-Fock-Bogoliubov or
Relativistic Mean Field (RMF) models \cite{LAL.99}), b) microscopic-macroscopic 
(Mic-Mac) models \cite{MOL.95,MOL.97,POM.03,MYE.96}

The Mic-Mac models are based on the well-known Strutinsky theorem.
According to this, the nuclear binding energy, hence the mass can be written
as sum of a smooth part, and an oscillatory part which  has its origins in the
quantum mechanical shell effects. The latter consists of the shell correction
energy and the pairing correlation energy which in 
the Mic-Mac models are evaluated in an external 
potential well. The smooth part is normally taken from
the liquid drop models of different degrees of sophistication. The largest
uncertainties arise in the calculation of shell corrections. The shell 
correction is calculated by taking the difference between the total quantum
mechanical energy of the given nucleus, and the corresponding `averaged' energy.
Usually, the averaging is achieved by the well-established Strutinsky scheme
\cite{STR.67,BUN.72}. This technique of calculating the averaged energies runs
into practical difficulties for finite potentials, since for carrying out the
Strutinsky averaging, one requires the discrete single-particle spectrum, with
cut-off well above (at least 3$\hbar\omega_0$, $\hbar\omega_0$ being the major
shell spacing) the Fermi energy. For a realistic potential, this condition is
not met, since continuum may start within $\sim\hbar\omega_0$ of the Fermi
energy. Standard practice is to discretise the continuum by diagonalising the
Hamiltonian in a basis of optimum size. A number of Mic-Mac calculations with
varying degree of success are available in the literature (see, for example,
\cite{MOL.95,MOL.97,POM.03,MYE.96}). The Mic-Mac models typically yield better
than $\sim$0.7 MeV rms deviation in the masses. All these models agree
reasonably well with each other and with experiment, but deviate widely 
among themselves
in the regions far away from the valley of stability.

The semi-classical Wigner-Kirkwood (WK) approach
\cite{WIG.32,KIR.33,JEN.75,RS.80,BRA.85,BRA.97,CEN.06,CEN.07}, on the other
hand, makes no explicit reference to the single-particle spectrum, and achieves
an accurate averaging of the given one-body Hamiltonian. Thus, the WK approach
is a good alternative to the conventional
Strutinsky smoothing scheme. The quantum mechanical energy is calculated by
diagonalising the one-body Hamiltonian in the axially symmetric deformed 
harmonic oscillator basis with 14 shells.
The difference between the total quantum mechanical energy and the WK energy
in the external potential well
yields the value of the shell correction for a given system. In the present
work, we propose to carry out a reliable microscopic-macroscopic calculation of
the nuclear binding energies (and hence the masses), employing the
semi-classical Wigner-Kirkwood (WK) $\hbar$ expansion
\cite{WIG.32,KIR.33,JEN.75,RS.80,BRA.85,BRA.97,CEN.06,CEN.07} for the
calculation of shell corrections instead of the Strutinsky scheme. 
An exploratory study of using the WK method to compute the smooth part 
of the energy has been reported earlier to test the validity of the
Strutinsky scheme, especially near the driplines \cite{NAZ.94}.

It is known that the WK level density ($g_{WK}(\varepsilon)$) with the 
$\hbar ^2$ correction term exhibits a $\varepsilon^{-1/2}$ divergence as
$\varepsilon\rightarrow0$, for potentials which vanish at large distances 
as for instance Woods-Saxon potentials (see, for example, Ref.\ \cite{SHL.91}). 
The Strutinsky level density, on the contrary, exhibits only a prominent peak 
as $\varepsilon\rightarrow0$. It was therefore concluded in Ref. \cite{VER.98} 
that the divergence of the WK level density as $\varepsilon\rightarrow0$ is 
unphysical, and the Strutinsky smoothed level density should be preferred. It 
should however be noted that the WK level densities, energy densities, etc., 
have to be understood in the mathematical sense of distributions and, 
consequently, only integrated quantities are meaningful. In fact, it has been 
shown \cite{CEN.07} that the integrated quantities such as the accumulated 
level densities are perfectly well behaved, even for $\varepsilon\rightarrow0$.

Pairing correlations are important for open shell nuclei. In the present work,
these are taken into account in the approximate particle number projected
Lipkin-Nogami scheme \cite{LIP.60,NOG.64,HCP.73}. Odd-even and
odd-odd nuclei are treated in an entirely microscopic fashion (odd nucleon
blocking method in the uniform filling approximation), allowing an improved 
determination of odd-even mass differences, see e.g. the discussion in \cite{XU.99}.
The majority of nuclei in the
nuclear chart are deformed. In particular, it is well known that inclusion of
deformation is important for reliable predictions of nuclear masses. Therefore,
here we incorporate in all three deformation degrees of freedom
($\beta_2,~\beta_4,~\gamma$). To our knowledge, no such detailed and extensive
calculation based on the WK method is available in the literature. 

The paper is organised as follows. We review the WK expansion in Section 2.
The choice of  the nuclear, spin-orbit, and Coulomb potentials forms the subject
matter of Section 3. Details of the WK calculations are discussed in Section 4. 
A systematic study of the WK energies for neutrons and protons as a function of
the deformation degrees of freedom is presented in Section 5. The shell corrections 
for the chains of Gd, Dy and Pb isotopes obtained by using our formalism 
are reported, and are compared with those calculated employing the traditional 
Strutinsky averaging technique, in Section 6. Section 7 contains
a brief discussion on the Lipkin-Nogami pairing scheme. As an illustrative
example, the calculation of the binding energies for selected 367 
spherical nuclei is presented and discussed in Section 8. 
Section 9 contains our summary and future outlook.
Supplementary material can be found in appendices A and B. 

\section{Semi-classical Wigner-Kirkwood Expansion}
Following Ref. \cite{JEN.75}, we consider
a system of $N$ non-interacting fermions at zero temperature.
Suppose that these fermions are moving in a given one-body potential including
the spin-orbit interaction. To determine 
the smooth part of the energy of such a system, we start with the quantal 
partition function for the  system:
\begin{eqnarray}
Z\left(\beta\right)~=~\mathrm{Tr}\left(\exp{(-\beta\hat{H})}\right) .
\label{Z_qm}
\end{eqnarray} 
Here, $\hat{H}$ is the Hamiltonian of the system, given by:
\begin{eqnarray} 
\hat{H}~=~\frac{-\hbar^2}{2m} \nabla^2 ~+~V(\vec{r})~+~\hat{V}_{LS}(\vec{r}) \,,
\end{eqnarray} 
where $V(\vec{r})$ is the one-body central potential and
$\hat{V}_{LS}(\vec{r})$ is the spin-orbit interaction.

In order to average out shell effects, the simplest one could do is replace 
the partition function in the above expression by the classical partition 
function. That is, one replaces the Hamiltonian in Eq. (\ref{Z_qm}) by the 
corresponding classical Hamiltonian. This yields the well-known Thomas-Fermi 
equations for particle number and energy. Way back in 1930's, E. Wigner 
\cite{WIG.32} and 
J. G. Kirkwood \cite{KIR.33} developed a systematic expansion of the partition 
function in powers of the Planck's constant, $\hbar$, its first term being the
classical partition function. Details of this method can be found in Refs.\
\cite{WIG.32,KIR.33,JEN.75,RS.80,BRA.85,BRA.97}. Such expansion of the quantal
partition function in powers of $\hbar$ is often known as Wigner-Kirkwood (WK) 
expansion. Systematic corrections to the Thomas-Fermi energy can be obtained by
using the WK expansion. 

In this work, we shall use the WK expansion up to fourth order. For brevity,
we represent the potentials and form factors without mentioning the dependence 
on the position vector. Ignoring the spin-orbit interaction, the WK expansion of
the partition function, correct up to fourth order is given by \cite{JEN.75}:
\begin{eqnarray} 
Z^{(4)}(\beta)&=&\frac{\beta^{-3/2}}{4 \pi^{3/2}}\left(\frac{2m}{\hbar^2}\right)^{3/2} 
\int d\vec{r} e^{-\beta V }
\left[1 - \frac{\beta^2 \hbar^2}{24m} \nabla^2 V \right. \nonumber \\
&& \left. + \frac{\beta^{3}}{1440}
\left(\frac{\hbar^2}{2m}\right)^{2}
\left\{-7\nabla^4V + 5\beta\left(\nabla^2V\right)^2+\beta\nabla^2\left(\nabla V\right)^2
\right\} \right] .
\label{Z_4}
\end{eqnarray} 
The spin-orbit interaction, in general, can be written as:
\begin{eqnarray} 
\hat{V}_{LS}~=~\frac{\iota\kappa\hbar^2}{2m}\left(\vec{\nabla}f \times \vec{\nabla} \right) \cdot \hat{\sigma}~,
\end{eqnarray} 
where $\hat{\sigma}$ is the unit Pauli matrix,
$\kappa$ is the strength of spin-orbit interaction, and $f$ is the spin-orbit
form factor. With the inclusion of such spin-orbit interaction, the 
WK expansion for the full partition function splits up into two parts:
\begin{eqnarray} 
Z_{WK}^{(4)}(\beta)~=~Z^{(4)}(\beta)~+~Z^{(4)}_{LS}(\beta)~.
\end{eqnarray} 
Here, $Z^{(4)}(\beta)$ is given by Eq. (\ref{Z_4}), and the spin-orbit
contribution to the partition function, correct up to fourth order in
$\hbar$, reads \cite{JEN.75}:
\begin{eqnarray} 
Z^{(4)}_{LS}&=&\frac{\kappa^2\beta^{1/2}}{8\pi^{3/2}}\left(\frac{2m}{\hbar^2}\right)^{1/2} 
             \int d\vec{r} e^{-\beta V} \left(\nabla f\right)^2  \nonumber \\
            &+& \frac{\beta^{1/2}}{96\pi^{3/2}}\left(\frac{\hbar^2}{2m}\right)^{1/2} \int d\vec{r} e^{-\beta V}
              \left[ \kappa^2 f_{2} - 2\kappa^3 f_{3} + 2 \kappa^4f_{4}\right] ,
\end{eqnarray} 
where
\begin{eqnarray} 
f_{2}&=&-\beta \left(\nabla f\right)^2 \left( \nabla^2 V\right) + \frac{1}{2}\nabla^2 \left(\nabla f\right)^2
- \left(\nabla^2 f\right)^2 + \nabla f \cdot \nabla \left(\nabla^2 f\right) \\
f_{3}&=&\left(\nabla f\right)^2 \nabla^2 f - \frac{1}{2} \nabla f \cdot \nabla \left(\nabla f\right)^2 \\
f_{4}&=&\left(\nabla f\right)^4 .
\end{eqnarray} 

The level density $g_{WK}$, particle number $N$, and energy $E$ can be
calculated directly from the WK partition function by Laplace inversion:
\begin{eqnarray} 
g_{WK}(\epsilon)~=~{\cal{L}}^{-1}_{\epsilon} Z_{WK}^{(4)}(\beta)~,
\end{eqnarray} 
\begin{eqnarray} 
N~=~{\cal{L}}^{-1}_{\lambda} \left( \frac {Z_{WK}^{(4)}(\beta)} {\beta} \right)
\label{N_qm}
\end{eqnarray} 
and
\begin{eqnarray} 
E~=~\lambda N~-~{\cal{L}}^{-1}_{\lambda} \left( \frac {Z_{WK}^{(4)}(\beta)} {\beta ^2} \right) ,
\label{E_qm}
\end{eqnarray} 
where $\lambda$ is the chemical potential, fixed by demanding the right 
particle number, and ${\cal{L}}^{-1}_{\lambda (\epsilon)}$ denotes the Laplace
inversion. Using the identity
\begin{eqnarray}
{\cal L}_{\lambda}^{-1} 
\left( \frac{e^{-\beta V}}{\beta^\mu} \right) ~=~
\frac{(\lambda - V)^{\mu-1}}{\Gamma(\mu)} \Theta(\lambda-V)~,\mathrm{for~}\mu~>0
\end{eqnarray} 
and  noting that, in order to get inverse Laplace transforms in convergent form,
\begin{eqnarray} 
e^{-\beta V}~=~\frac{-1}{\beta}\frac{\partial e^{-\beta V}}{\partial V}~,
\label{trick}
\end{eqnarray} 
one obtains the level 
density for each kind of nucleons assuming spin degeneracy:
\begin{eqnarray} 
g_{WK}(\epsilon)&=& \frac{1}{3\pi^2}\left( \frac{2m}{\hbar^2} \right)^{3/2} \int
d\vec{r} \left[\frac{3}{2}\left(\epsilon - V\right)^{1/2} 
 + \frac{\hbar^2}{4m}\left\{\frac{3}{4}\kappa^2\left(\nabla f\right)^2 \left(\epsilon - V\right)^{-1/2}
 \right.\right. \nonumber \\
 && \left.\left. \hskip 5.95cm + \frac{1}{16}\Delta V \left(\epsilon - V\right)^{-3/2}\right\} \right] 
\Theta \left(\epsilon-V\right) ,
\label{gWK}
\end{eqnarray} 
the particle number:
\begin{eqnarray} 
N &=& \frac{1}{3\pi^2}\left(\frac{2m}{\hbar^2}\right)^{3/2} \int d\vec{r} 
\left[\left(\lambda-V\right)^{3/2} 
        - \frac{\hbar^2}{32m}\left(\lambda-V\right)^{-1/2} \nabla^2 V \right. \nonumber \\ 
       && \hspace{5.7cm}\left. + \frac{3\hbar^2\kappa^2}{8m}\left(\lambda-V\right)^{1/2} \left(\nabla f\right)^2\right]
\Theta \left(\lambda-V\right) ,
\label{N_WK}
\end{eqnarray} 
and the energy:
\begin{eqnarray} 
E = \lambda N &-& \frac{1}{3\pi^2}\left(\frac{2m}{\hbar^2}\right)^{3/2} \int d\vec{r} \left [
                 \frac{2}{5}\left(\lambda-V\right)^{5/2} - \frac{\hbar^2}{16m} \left(\lambda-V\right)^{1/2} \nabla^2 V \right]
      \Theta \left(\lambda-V\right) \nonumber \\
  &-& \frac{1}{5760\pi^2} \left(\frac{\hbar^2}{2m}\right)^{1/2} \left[ 
\int d\vec{r} \left(\lambda-V\right)^{-1/2}\left\{ -7\nabla^4 V\right\}  \right. \nonumber \\
  && \left. \hspace{3.65cm} 
    - \frac{1}{2} \int d\vec{r} \left(\lambda-V\right)^{-3/2} \left\{ 5\left(\nabla^2V\right)^2+\nabla^2\left(\nabla V\right)^2 \right\}
          \right] \Theta \left(\lambda-V\right) \nonumber \\
  &-&\frac{\kappa^2}{6\pi^2}\left(\frac{2m}{\hbar^2}\right)^{1/2} \int d\vec{r} \left(\lambda-V\right)^{3/2} \left(\nabla f\right)^2 
    \Theta \left(\lambda-V\right)  \nonumber \\
  &-& \frac{1}{48\pi^2}\left(\frac{\hbar^2}{2m}\right)^{1/2} \int d\vec{r} \left(\lambda-V\right)^{1/2} 
\left[ \kappa^2 \left\{ \frac{1}{2}\nabla^2 \left( \nabla f\right)^2 - \left(\nabla^2 f\right)^2 
         + \nabla f \cdot \nabla \left(\nabla^2 f\right) \right. \right. \nonumber \\ 
  & & \left.\left. - \frac { \left(\nabla f\right)^2 \nabla^2 V } { 2 \left(\lambda-V\right) } \right\} 
      - 2\kappa^3 \left\{ \left( \nabla f\right)^2 \nabla^2 f - \frac{1}{2} \nabla f \cdot \nabla \left(\nabla f\right)^2 \right\}
      +2\kappa^4 \left( \nabla f\right)^4
\right ] \Theta \left(\lambda-V\right) \nonumber \\
\label{E_WK}
\end{eqnarray} 
It should be noted that we have explicitly assumed that all the derivatives of
the potential $V$ and the spin-orbit form factor $f$ exist. The expansion
defined here is therefore not valid for potentials with sharp surfaces. This
automatically puts a restriction on the choice of the Coulomb potential: the
conventional uniform distribution approximation for the charge distribution
cannot be used in the present case. We shall discuss this point at a greater
length in the next section. The integrals in the above expressions are cut off
at the turning points, defined via the step function. The chemical potential
$\lambda$ appearing in these equations is determined from Eq. (\ref{N_WK}),
separately for neutrons and protons. Further, it is interesting to note that the
spin-orbit contribution to the particle number $N$ as well as to the energy $E$
appears only in the second order in $\hbar$. Secondly, the level density and
particle number are calculated only up to the order $\hbar^2$. It can be shown
\cite{JEN.75} that for the expansion correct up to fourth order in $\hbar$, it
is sufficient to take $Z_{WK}^{(4)}$ up to order $\hbar^2$ in Eq.
(\ref{N_qm}) to find the chemical potential (and hence the particle number),
whereas one has to take the full partition function $Z_{WK}^{(4)}$  up to order
$\hbar^4$ in Eq. (\ref{E_qm}) to compute the energy in the WK approach.

The divergent terms appearing in Eq. (\ref{E_WK}) are treated by differentiation 
with respect to the chemical potential. Explicitly:
\begin{eqnarray} 
\partial_{\lambda} \left( \lambda - V \right)^{1/2} &=& \frac{1}{2}\left( \lambda - V \right)^{-1/2}\\
\partial^{2}_{\lambda} \left(\lambda - V \right)^{1/2} &=& -\frac{1}{4}\left(\lambda - V\right)^{-3/2}
\end{eqnarray} 
In practice, the differentiation with 
respect to chemical potential is carried out {\it after} evaluation of the 
relevant integrals. Numerically, this approach is found to be stable. Its reliability has
been checked explicitly by reproducing the values of fourth-order WK corrections 
quoted in Ref.\ \cite{JEN.75}.

The WK expansion thus defined, converges very rapidly for the harmonic
oscillator potential: the second-order expansion itself is enough for most
practical purposes. The convergence for the Woods-Saxon potential, however, is
slower than that for the harmonic oscillator potential, but it is adequate \cite{JEN.75a}.
For example, for $\sim 126$ particles, the Thomas-Fermi energy is typically of the
order of $10^3$ MeV, the second-order ($\hbar^2$) correction contributes a few
10's of MeVs, and the fourth-order ($\hbar^4$) correction yields a contribution
of the order of 1 MeV. This point will be discussed in greater details later.
It is also important to note that the WK $\hbar$ expansion of the 
density matrix has a variational character and that a variational theory based on a 
strict expansion of the of $\hbar$ has been established 
\cite{PS.93}.

The WK approach presented here should be distinguished from the
extended Thomas-Fermi (ETF) approach. Divergence problems at the classical
turning points (see the particle number and energy expressions above) can be
eliminated by expressing the kinetic energy density as a functional of the local
density. This is achieved by eliminating the chemical potential, the local potential,
and the derivatives of the local potential (for further details, see Ref. \cite{CEN.98}). 
It cannot be accomplished in closed form, and has to be done iteratively, 
leading to a functional series
for the kinetic energy density. The resulting model is what is often referred to
as the ETF approach. The WK approach as presented here, in this sense, is the
starting point for ETF approach (further details of ETF can be found in Refs.\
\cite{GRA.79,GRA.80,BRA.85,CEN.90,BRA.97,CEN.07}). The conventional ETF
approach exhibits somewhat slower convergence properties which has been
attributed to a non-optimal sorting out of terms of each 
given power in $\hbar$ \cite{CEN.07,CEN.98}. 

\section{Choice of potential}
\subsection{Form of the Nuclear Potential}
The spherically symmetric nuclear mean field is well represented by
the Woods-Saxon (WS) form \cite{WS.54}, given by:
\begin{eqnarray} 
V(r)~=~\frac{V_0}{1 + \exp{((r~-~R_0)/a)}}~,
\label{ws0}
\end{eqnarray} 
where $V_0$ is the strength of the potential, $R_0$ is the half-density
radius, and $a$ is the diffuseness parameter. The WS form factor defined here,
can be easily generalised to take the deformation effects into account. Note 
that the distance function $l(r)=r-R_0$ appearing in Eq. (\ref{ws0}) can be 
interpreted as the minimum distance of a given point to the nuclear surface,
defined by $r~=~R_0$. One might thus generalise it to the case of deformed 
surfaces as well. Using the standard expansion in terms of spherical harmonics, 
a general deformed surface may be defined by the relation $r~=~r_s$, where
\begin{eqnarray} 
r_s~=~CR_0(1~+~\sum_{\lambda,\mu}\alpha_{\lambda,\mu}Y_{\lambda,\mu})~.
\label{def}
\end{eqnarray} 
Here, the $Y_{\lambda,\mu}$ functions are the usual spherical harmonics and the
constant $C$ is the volume conservation factor (the volume enclosed by the
deformed surface should be equal to the volume enclosed by an equivalent
spherical surface of radius $R_0$):
\begin{eqnarray} 
C~=~\left[ \frac{1}{4\pi}\int_{\Omega}\left\{1~+~\sum_{\lambda,\mu}\alpha_{\lambda,\mu}
Y_{\lambda,\mu}(\Omega)\right\}^3 d\Omega \right]^{-1/3} .
\end{eqnarray} 

The distance function to be used in the WS potential would be the minimum
distance of a given point to the nuclear surface defined by $r~=~r_s$. Such
definition has been used quite extensively in the literature, with good success
(see, for example, Refs.\ \cite{DUD.79,DUD.81,NAZ.85,CWI.87,RW.91}). However, in
the present case, this definition is not convenient, since it should be noted
that the calculation of this distance function involves the minimisation of a
segment from the given point to the nuclear surface. This in turn implies that
each calculation of the distance function (for given $r$, $\theta$, and $\phi$
coordinates: we are assuming a spherical polar coordinate system here) involves
the calculation of two surface angles $\theta_s$ and $\phi_s$, and these are 
implicit functions of $r$, $\theta$, and $\phi$. See Fig. (8) in Appendix A for 
details. Since the WK calculations
involve differentiation of the WS function, one also needs to differentiate 
$\theta_s$ and $\phi_s$, which are implicit functions of $r$, $\theta$, and
$\phi$. 

Alternatively, the distance function for the deformed Woods-Saxon potential can
be written down by demanding that the rate of change of the potential calculated
normal to the nuclear surface and evaluated at the nuclear surface should be a 
constant \cite{DAM.69} which, indeed, is the case for the spherical Woods-Saxon
form factor. Thus,
\begin{eqnarray} 
\left\{\hat{n}\cdot\nabla{V\left(\vec{r}\right)}\right\}_{r=r_s}
~=~\mathrm{constant} \,,
\label{cond1}
\end{eqnarray} 
where $\hat{n}$ is the unit vector normal to the surface ($r~=~r_s$) and is
given by
\begin{eqnarray} 
\hat{n}~=~\frac{\nabla {(r~-~r_s)}}{|\nabla {(r~-~r_s)}|} \,.
\end{eqnarray} 
In fact, the above condition (\ref{cond1}) is related to the observation that 
the second derivative of the spherical Woods-Saxon form factor vanishes at 
the nuclear surface, defined by $r~=~R_0$. The resulting distance function 
is given by \cite{BEN.68}:
\begin{eqnarray} 
l(\vec{r})~=~\frac{r~-~r_s}{|\nabla{(r~-~r_s)}|_{r=r_s}}~,
\label{dist}
\end{eqnarray}
where $r_s$ is as defined in Eq. (\ref{def}). The denominator is 
evaluated at $r=r_s$. Writing the $\theta$ and $\phi$
derivatives of $r_s$ as $A$ and $B$ respectively, we get:
\begin{eqnarray} 
l(\vec{r})~=~\frac{(r~-~r_s)}{\sqrt{1~+~\gamma^2/{r_s}^2}}~,
\end{eqnarray} 
with
\begin{eqnarray} 
\gamma^2~=~ A^2~+~B^2\csc^2\theta~.
\end{eqnarray} 
In the present work, we use the distance function as defined in 
Eq. (\ref{dist}). The WS potential thus reads
\begin{eqnarray} 
V\left(\vec{r}\right)~=~\frac{V_0}{1 + \exp{(l(\vec{r})/a)}}~.
\label{ws1}
\end{eqnarray} 
It is straightforward to check that the Woods-Saxon
potential defined with the distance function as given by Eq. (\ref{dist}) 
satisfies the condition (\ref{cond1}). 
Substituting this Woods-Saxon potential 
in $\hat{n}\cdot\nabla{V\left(\vec{r}\right)}$, we get
\begin{eqnarray} 
\hat{n}\cdot\nabla{V\left(\vec{r}\right)}&=& \frac{V_0}{a}f(\vec{r})\left( f(\vec{r}) - 1\right)\hat{n}\cdot \nabla l(\vec{r}) \nonumber \\
                                         &=& \frac{V_0}{a}f(\vec{r})\left( f(\vec{r}) - 1\right)
  \left[ \frac{| \nabla \left(r - r_s\right) |}{| \nabla \left(r - r_s\right) |_{r=r_s}} \right. \nonumber \\
    &&  \hspace{4cm} \left. + \left(r - r_s\right)\frac{ \nabla \left(r - r_s\right) }{| \nabla \left(r - r_s\right) |}
                             \cdot \nabla \frac{1}{| \nabla \left(r - r_s\right) |_{r=r_s}} \right] .
\label{cond2}
\end{eqnarray} 
Here, $f(\vec{r}) = [1 + \exp{(l(\vec{r})/a)}]^{-1}$ is the Woods-Saxon form 
factor. Clearly, at the surface defined by $r = r_s$, 
the quantity $\hat{n}\cdot\nabla{V\left(\vec{r}\right)}$ is constant.

\subsection{Deformation Parameters}
In practice, we consider three deformation degrees of freedom, 
namely, $\beta_2$, $\beta_4$ and $\gamma$. These parameters 
are related with the parameters $\alpha_{\lambda,\mu}$ introduced in 
Eq. (\ref{def}). Note that for the given choice of deformation
parameters, $\lambda$ takes values 2 and 4. The projection
$\mu$ takes the values 0, $\pm2$ for $\lambda=2$ and the values
0, $\pm2$ and $\pm4$ for $\lambda=4$. Further, existence of 
symmetry planes $(x,y)$, $(y,z)$ and $(z,x)$ implies that \cite{NAZ.85}
\begin{eqnarray} 
\alpha_{2,2}~=~\alpha_{2,-2},~
\alpha_{4,2}~=~\alpha_{4,-2},~
\alpha_{4,4}~=~\alpha_{4,-4} \,. \nonumber
\end{eqnarray} 
Thus, we get:
\begin{eqnarray} 
r_s\left(\theta,\phi\right)&=&CR_0 \left[ 1 + \alpha_{2,0}Y_{2,0}\left(\theta\right)
     + \alpha_{2,2}\left\{Y_{2,2}\left(\theta,\phi\right) + Y_{2,-2}\left(\theta,\phi\right)\right\} 
     + \alpha_{4,0}Y_{4,0}\left(\theta\right) \right. \nonumber \\
    &+& \left. 
       \alpha_{4,2}\left\{Y_{4,2}\left(\theta,\phi\right) + Y_{4,-2}\left(\theta,\phi\right)\right\} 
     + \alpha_{4,4}\left\{Y_{4,4}\left(\theta,\phi\right) + Y_{4,-4}\left(\theta,\phi\right)\right\} \right] ,
\end{eqnarray} 
where
\begin{eqnarray} 
\alpha_{2,0}&=&\beta_{2} \cos \gamma \\
\alpha_{2,2}&=&-\sqrt{\frac{1}{2}}\beta_{2} \sin \gamma \\
\alpha_{4,0}&=&\frac{1}{6}\beta_{4}\left(5\cos^2\gamma + 1\right) \\
\alpha_{4,2}&=&-\sqrt{\frac{30}{144}}\beta_{4}\sin 2\gamma \\
\alpha_{4,4}&=&\sqrt{\frac{70}{144}}\beta_{4}\sin^2\gamma \,.
\end{eqnarray} 
For further details, see Ref. \cite{NAZ.85}.

\subsection{Woods-Saxon Parameters}
The parameters \cite{RW.99} appearing in the Woods-Saxon potential are as 
defined below:
\begin{enumerate}
\item Central potential: \\
\noindent 
a. Strength:
\begin{eqnarray} 
V_0~=~-U_0 \left\{ 1 \mp U_1 \frac{N-Z}{A} \right\}
\end{eqnarray} 
with $U_0$=53.754 MeV and $U_1$=0.791.

\noindent 
b. Half-density radius:
\begin{eqnarray} 
R_0~=~r_0A^{1/3}\left\{ 1 \mp c_1 \frac{N-Z}{A} \right\} + c_2
\end{eqnarray} 
with $r_0$=1.19 fm, $c_1$=0.116 and $c_2$=0.235 fm.

\noindent 
c. Diffuseness parameter: assumed to be same for neutrons and protons,
and has the value $a~=~0.637$ fm.

\item Spin-orbit potential: \\
\noindent 
a. Strength:
\begin{eqnarray} 
V_{SO}~=~\lambda_0 U_0 \frac{\hbar^2}{4m^2}\left\{ 1 \mp U_2 \frac{N-Z}{A} \right\}
\end{eqnarray} 
with $U_0$=53.754 MeV, $\lambda_0$=29.494 and $U_2$=0.162.

\noindent 
b. Half density radius and diffuseness parameter are taken to be the same as 
those for the central potential.
\end{enumerate}

In these expressions, the $+$ ($-$) sign holds for protons (neutrons).

The parameters have the isospin dependence of the central and spin-orbit 
potentials ``built-in''. This potential yields a reasonably good description 
of charge radii (both magnitude and isospin dependence) as well as of moments of 
inertia for a wide range of nuclei. It has been used extensively in the 
total Routhian surface (TRS) calculations, and it has been quite successful 
in accurately reproducing energies of single-particle as well as collective 
states \cite{RW.05}.

\subsection{Coulomb potential}
The Coulomb potential is calculated by folding the point proton density
distribution  $\rho(\vec{r}')$, assumed to be of Woods-Saxon form. For simplicity, 
its parameters are assumed to be the same as those for the nuclear potential 
of protons. The reason for using folded potential here is, as we 
have indicated in section II, the WK expansion is not valid 
for potentials with sharp surfaces.

The Coulomb potential for the
extended charge distribution is given by:
\begin{eqnarray} 
V_C(\vec{r})~=~e^2\int \rho(\vec{r}') \frac {1} {|\vec{r}-\vec{r}'|} d\vec{r}'~.
\label{coulpot}
\end{eqnarray} 
Here, 
\begin{eqnarray} 
|\vec{r}-\vec{r}'|~=~\left\{ r^2 + r'^2 - 2 rr' \cos\Psi \right\}^{1/2}~,
\end{eqnarray} 
where
\begin{eqnarray} 
\cos\Psi~=~\cos\theta\cos\theta' + \sin\theta\sin\theta'\cos(\phi-\phi')~,
\end{eqnarray} 
as explained in Appendix A. 

\begin{figure}[htb]
\centerline{\epsfig{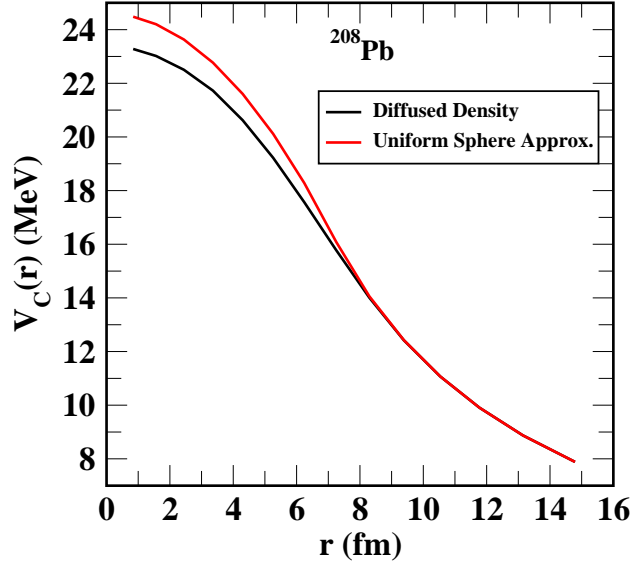}}
\caption{Coulomb potentials obtained by using diffuse density and 
sharp surface approximation for $^{208}$Pb.}
\label{pot}
\end{figure}

It is instructive at this point, to compare the Coulomb potential calculated 
from the diffuse density with the corresponding potential obtained by using the
conventional uniform density (sharp surface) approximation. Such comparison for
$^{208}$Pb is plotted in Fig.~\ref{pot}. The radius parameter for the diffuse
density approach as well as for the sharp surface approximation is assumed to be
equal to 7.11 fm (see the discussion on the choice of the Woods-Saxon parameters
in Section 3). It can be seen that in the exterior region, the two potentials
agree almost exactly, as expected. In the interior, however, the potential obtained
from the diffuse density turns out to be somewhat less repulsive than that from the 
density with sharp surface.

\section{Details of the WK Calculations}
In the present work, we restrict our calculations to three deformation
degrees of freedom, namely, $\beta_2$, $\beta_4$ and the angle $\gamma$. 
The inclusion of $\gamma$ allows to incorporate triaxiality. Thus, 
the present WK calculation is genuinely three dimensional. In principle
it is natural to use a cylindrical coordinate system here. The
spherical polar coordinates, however, turn out to be more convenient. 
The reason is, the cylindrical coordinates involve two length variables, 
and one angular coordinate which means that the turning points have to 
be evaluated for two coordinates ($\rho$ and $z$). This makes the calculations
very complicated. On the other hand, the spherical polar coordinates
involve only one length variable, and thus the turning points are 
to be evaluated only for one coordinate ($r$). The numerical 
integrals involved are evaluated using Gaussian quadrature.

The first step in the WK calculations is the determination of the chemical
potential. This has to be done iteratively, using Eq. (\ref{N_WK}). Since the 
turning points are determined by the chemical potential, they have to be
calculated using a suitable numerical technique at each step. 
Once the values of the chemical potential are known, the WK energies
up to second order can be calculated in a straightforward way.
The fourth-order calculations are very complicated, since they
require higher-order derivatives of nuclear potentials, spin-orbit
form factors, and the Coulomb potential. The former can be evaluated
analytically in the present case. The expressions are extremely 
lengthy, and we do not present them here. Comparatively, the 
derivatives of the Coulomb potential look simple; the Laplacian and
Laplacian of Laplacian are completely straightforward: the 
former is proportional to the proton density and the latter is
just the Laplacian of the WS form factor. However, the calculations
also need terms like Laplacian of the gradient squared of the total potential.
In the case of protons, this involves one crossed term:  
\begin{eqnarray} 
\nabla^2 \left(\nabla V_C(\vec{r}) \cdot \nabla V_N(\vec{r}) \right) ,
\label{cross}
\end{eqnarray} 
where $V_C$ is the Coulomb potential and $V_N$ is the nuclear potential.
The determination of such objects is tricky. It turns out that if one
uses the form of the Coulomb potential defined above, the calculation of
expression (\ref{cross}) becomes numerically unstable. 

There exists an alternative for of the Coulomb potential:
\begin{eqnarray} 
V_C(\vec{r})~=~\frac{e^2}{2}\int d\vec{r}' |\vec{r}-\vec{r}'| \nabla^{2}_{\vec{r}'} \rho(\vec{r}')~,
\label{altcoul}
\end{eqnarray} 
where the notation $\nabla^{2}_{\vec{r}'}$ means that the Laplacian is 
calculated with respect to the variables $r'$, $\theta '$, and $\phi '$.
Eqs. (\ref{coulpot}) and (\ref{altcoul}) are exactly equivalent. This is 
proved explicitly in Appendix B. With this form, one can calculate the
first and second derivatives ({\it not} the Laplacian) of the Coulomb potential.
Calculation of the higher-order derivatives of the Coulomb potential, even
with the form defined in Eq. (\ref{altcoul}), turns out to be numerically
unstable. For this purpose, we employ the Poisson's equation. The details 
are presented in Appendix B. Once all the derivatives are known, the 
fourth-order WK calculations can be carried out. 

It turns out that the WK calculations for protons are very time consuming. This
is due to the fact that the calculation of Coulomb potential (Eq. (\ref{coulpot})), 
in general, involves evaluation of three dimensional integral for {\it each} point
$(r,\theta,\phi)$. Typically, it takes few tens of minutes to 
complete one such calculation. This is certainly not desirable, since our aim 
is to calculate the masses of the nuclei spanning the entire periodic table. To speed 
up the calculations, we use the well-known technique of interpolation. 
Since we are using spherical polar coordinates, the turning points
are to be evaluated only for the radial coordinate, $r$. For the {\it entire} 
WK calculation, the $\theta$ and $\phi$ mesh points remain the same 
(over the domains $[0,\pi]$ and $[0,2\pi]$, respectively), whereas the $r$ 
mesh points change from step to step. This happens in particular during the 
evaluation of the chemical potential. Once the convergence of the particle 
number equation (Eq. (\ref{N_WK})) is achieved, the $r$ mesh points as well, 
remain fixed.

Motivated by the above observations, we apply the following procedure:
\begin{enumerate}
\item Before entering into the actual WK calculations (determination 
of chemical potential, etc.), for {\it each pair} of $\theta$ and $\phi$ mesh
points, we calculate the Coulomb potential (Eq. (\ref{coulpot})) over a range 0 to 16 fm at
equidistant radial mesh points (the typical mesh size being 0.1 fm).
\item Next, for each pair of $\theta$ and $\phi$ mesh points, we fit a 
polynomial of degree $9$ in the radial coordinate $r$ to the Coulomb potential
calculated in the above step. 
Thus, the fitting procedure is to be repeated $N_\theta \times N_\phi$
times, $N_\theta$ ($N_\phi$) being total number of mesh points for the
$\theta$ ($\phi$) integration. 
\end{enumerate}
Thus, for any given value of radial the coordinate $r$ (and fixed $\theta$ and
$\phi$), the Coulomb potential can be easily calculated just by evaluating the
9$^{th}$ degree polynomial in $r$. It is found that this interpolation procedure
is very accurate. The maximum percentage difference between the fitted and 
the exact Coulomb potentials is 0.4\% for a highly deformed nucleus.

\section{Variation of Wigner-Kirkwood energies with deformation parameters}
\begin{figure}[htb]
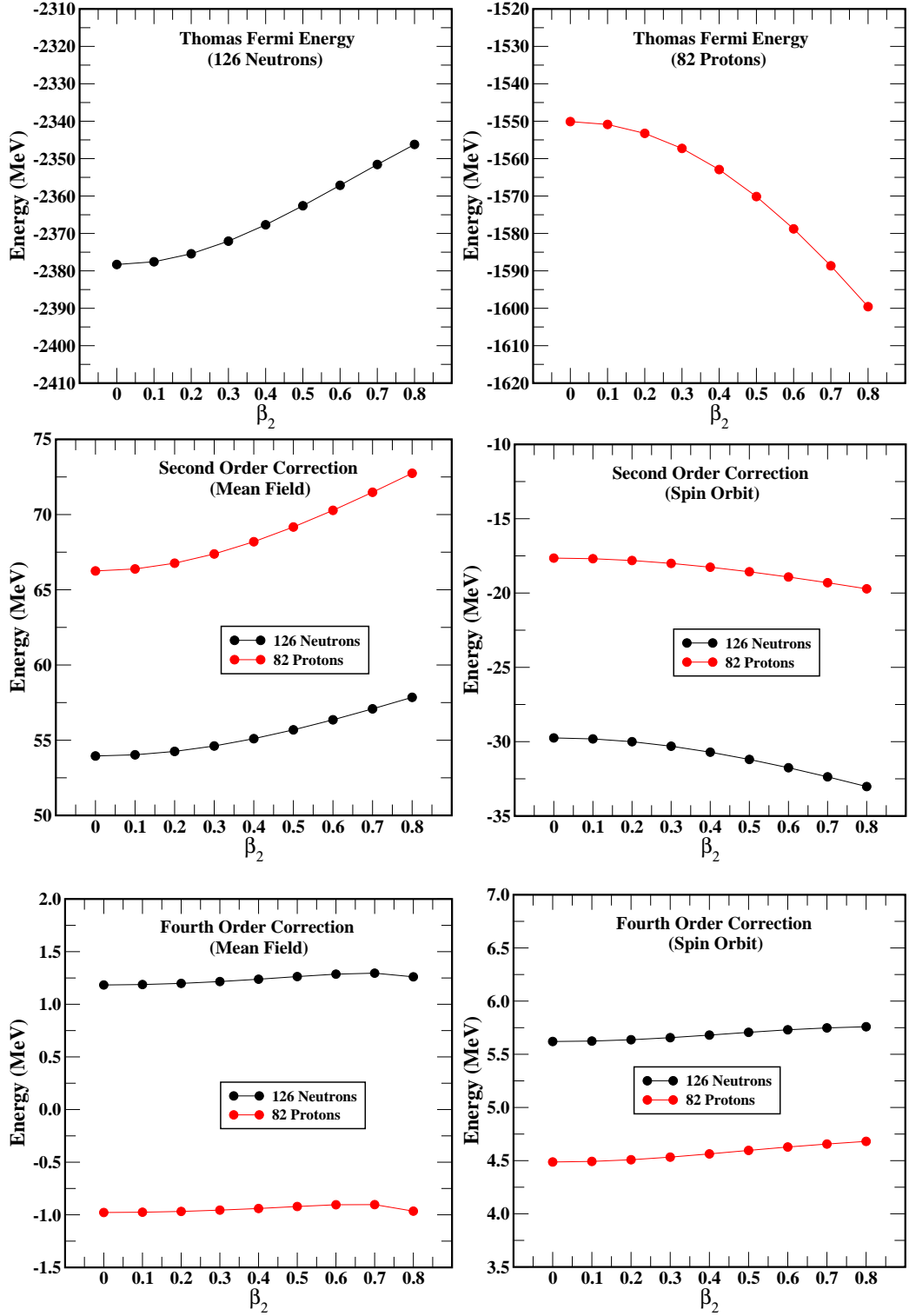

\begin{minipage}{0.92\textwidth}
\centerline{\hbox{ \epsfig{file=b2_tf_n.eps,width=0.45\textwidth} \hfill
                   \epsfig{file=b2_tf_p.eps,width=0.45\textwidth}}}
\end{minipage}
\vspace{10pt}
\begin{minipage}{0.92\textwidth}
\centerline{\hbox{ \epsfig{file=b2_hb2.eps,width=0.45\textwidth} \hfill  
                   \epsfig{file=b2_hb2s.eps,width=0.45\textwidth}}}
\end{minipage}
\vspace{10pt}
\begin{minipage}{0.92\textwidth}
\centerline{\hbox{ \epsfig{file=b2_hb4.eps,width=0.45\textwidth} \hfill  
                   \epsfig{file=b2_hb4s.eps,width=0.45\textwidth}}}
\end{minipage}
\caption{Wigner Kirkwood energies for 126 neutrons and 82 protons as 
a function of $\beta_2$. Here,  $\beta_4=0$ and $\gamma=0$. Thomas
Fermi energies, second-order corrections and the fourth-order 
corrections are shown in upper, middle and bottom panels respectively.}
\label{figt1}
\end{figure}

\begin{figure}[htb]
\begin{minipage}{0.92\textwidth}
\centerline{\hbox{ \epsfig{file=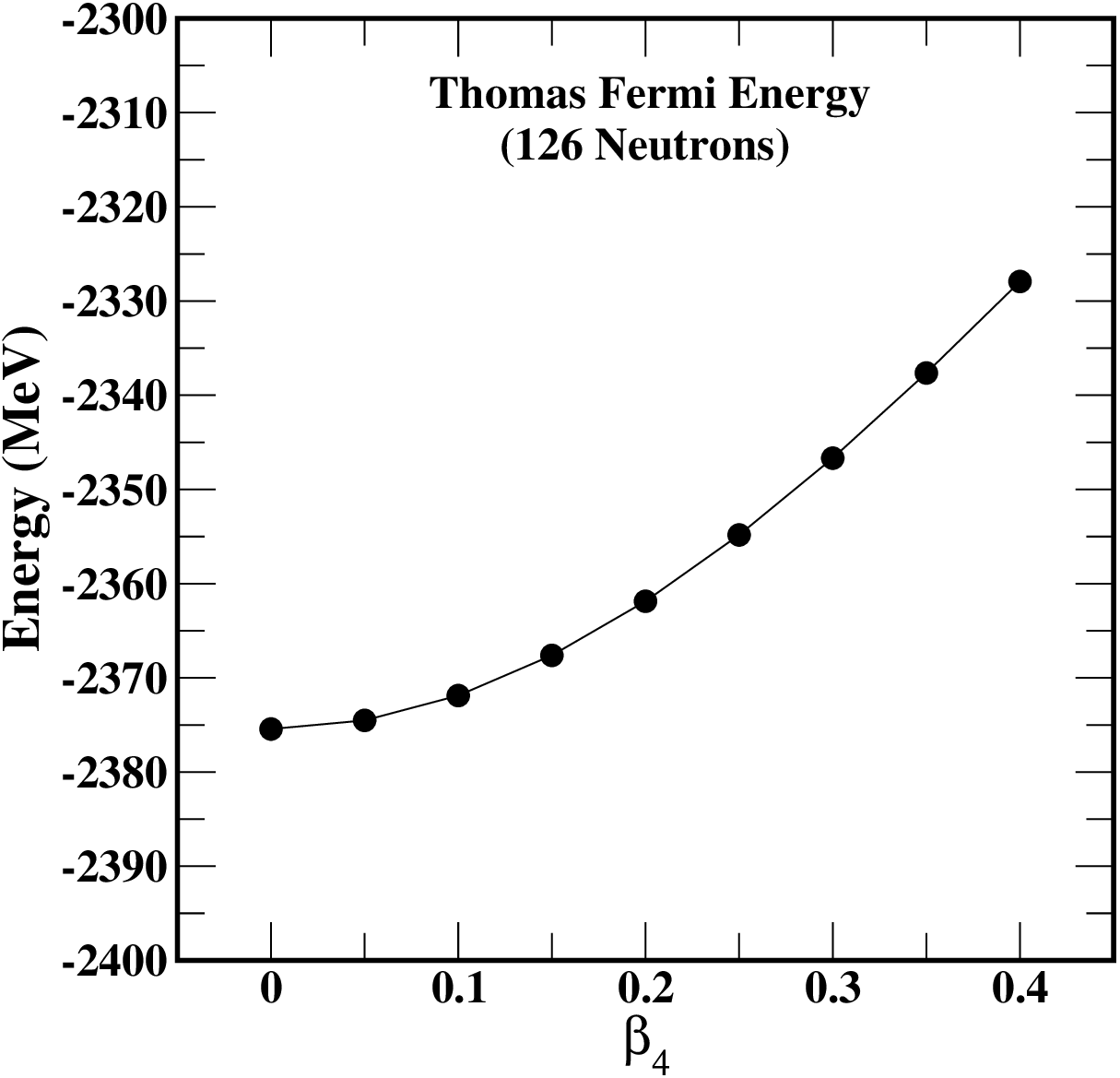,width=0.45\textwidth} \hfill
                   \epsfig{file=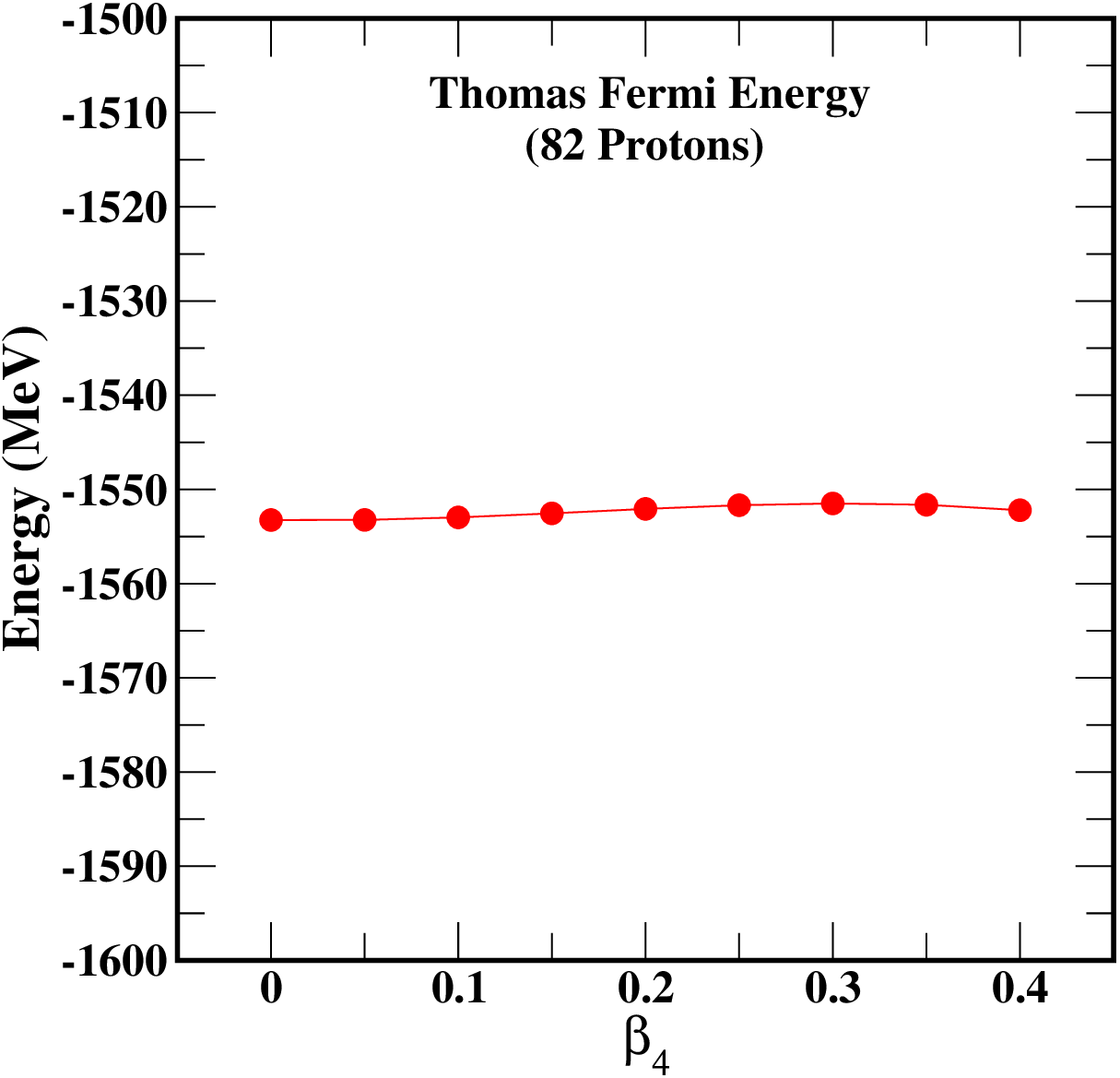,width=0.45\textwidth}}}
\end{minipage}
\vspace{10pt}
\begin{minipage}{0.92\textwidth}
\centerline{\hbox{ \epsfig{file=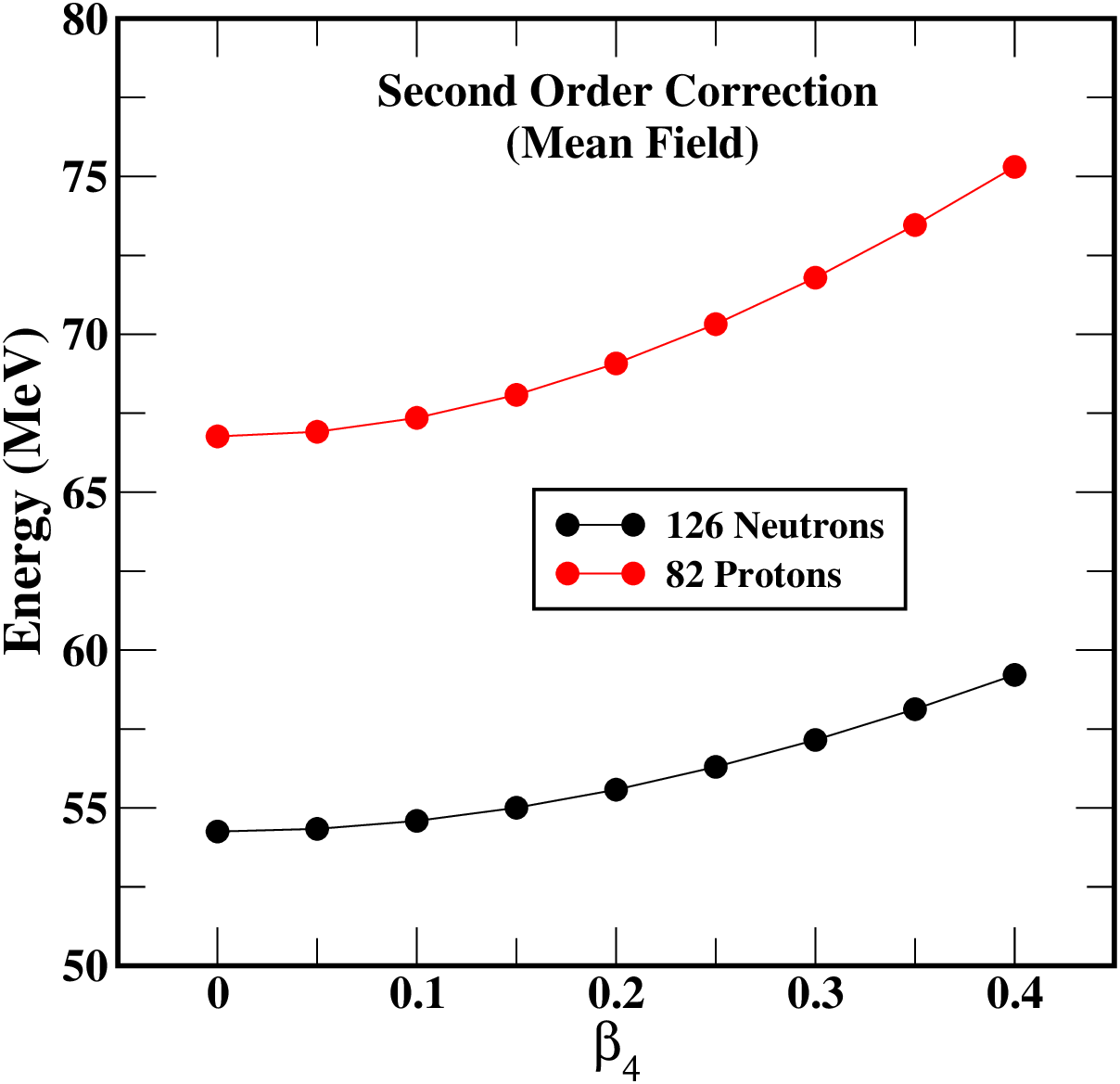,width=0.45\textwidth} \hfill  
                   \epsfig{file=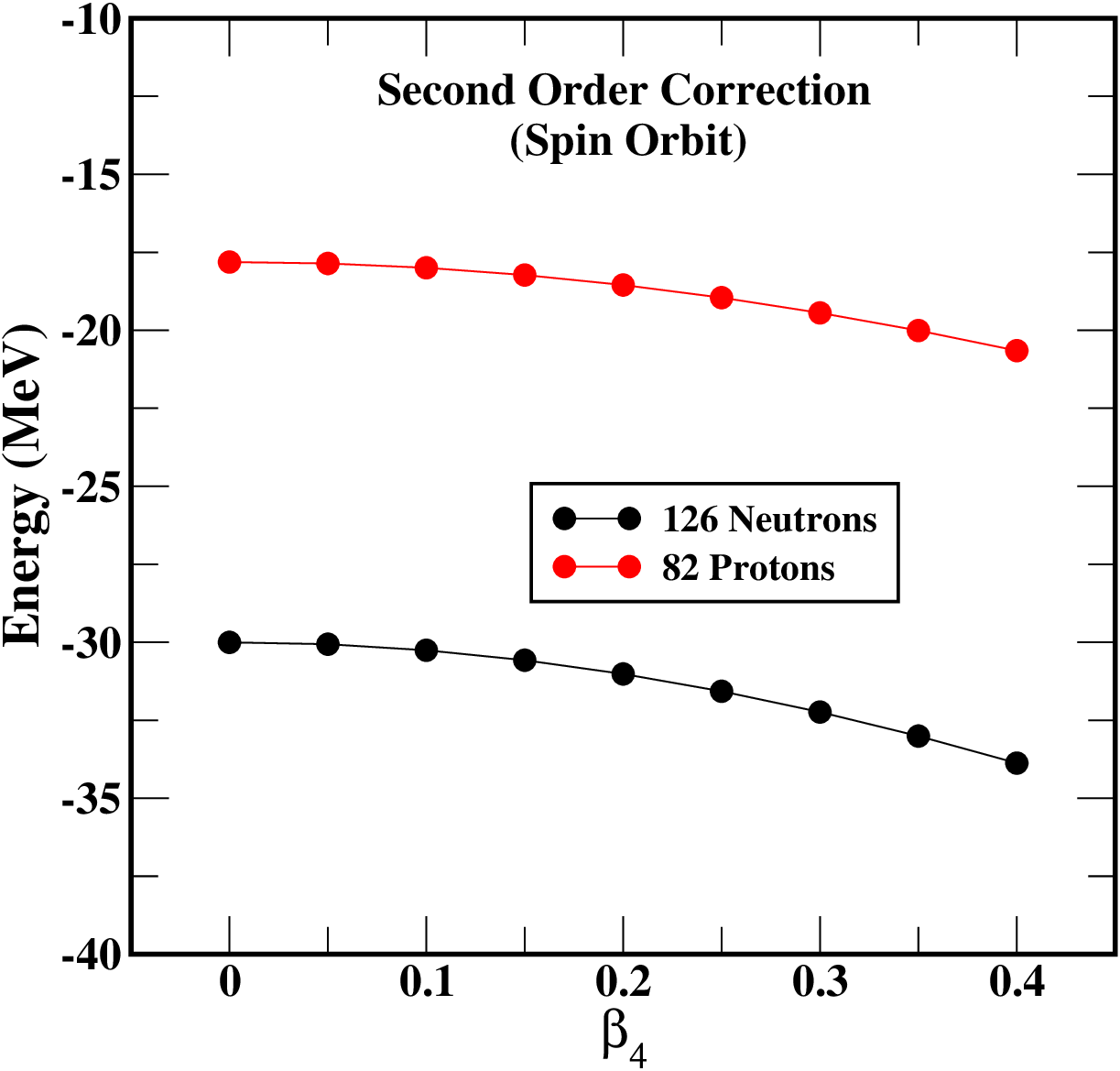,width=0.45\textwidth}}}
\end{minipage}
\vspace{10pt}
\begin{minipage}{0.92\textwidth}
\centerline{\hbox{ \epsfig{file=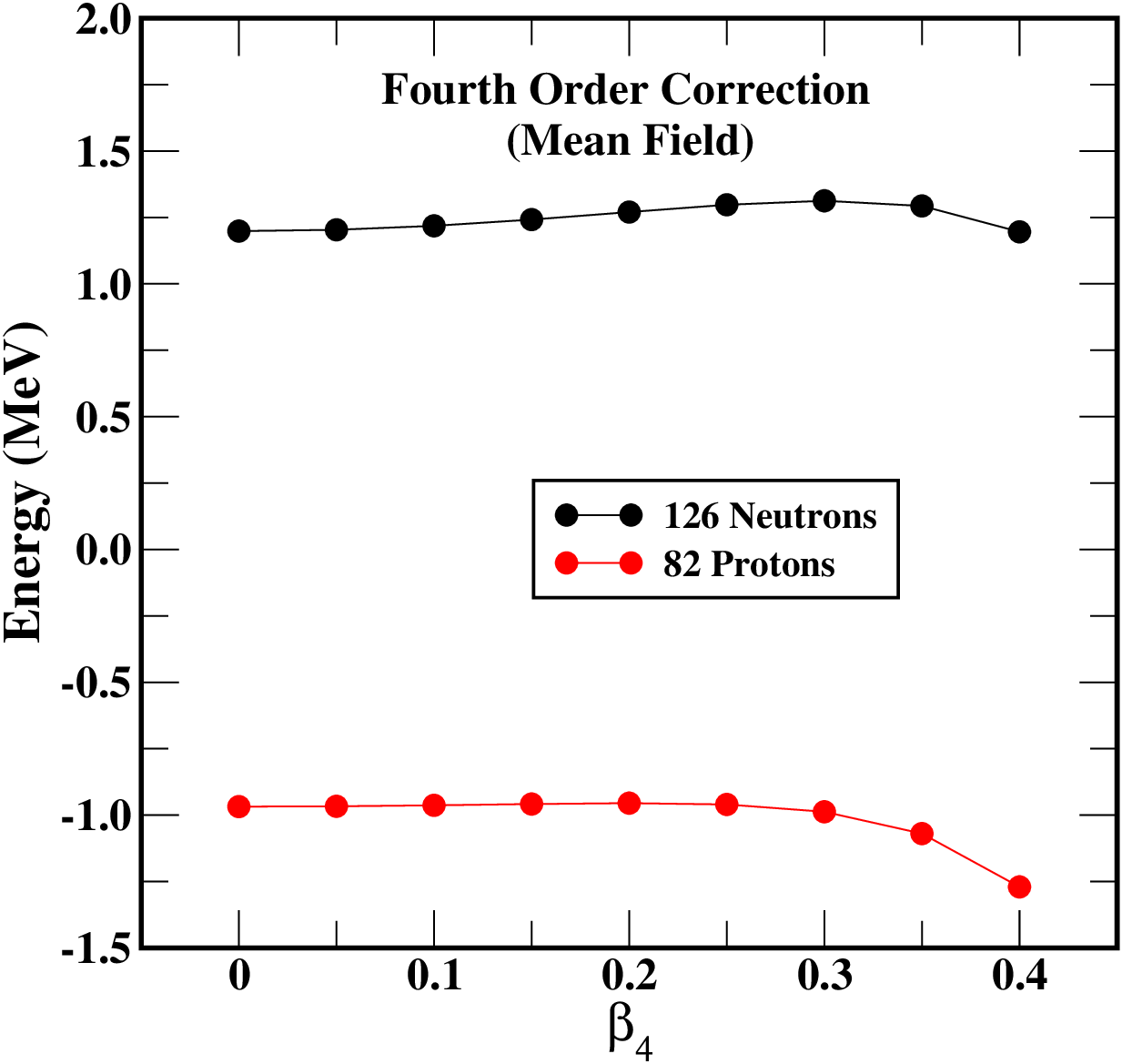,width=0.45\textwidth} \hfill  
                   \epsfig{file=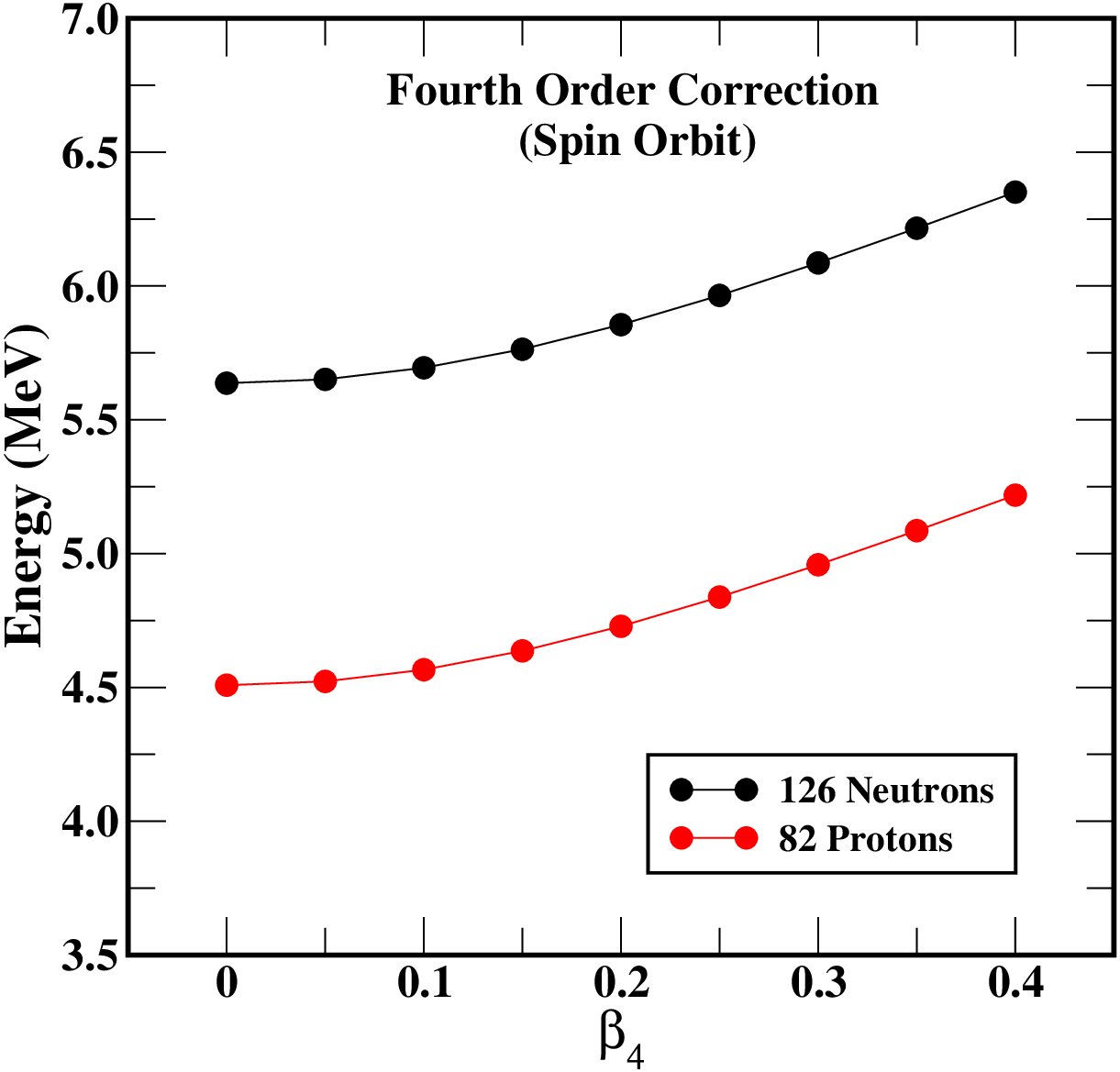,width=0.45\textwidth}}}
\end{minipage}
\caption{Wigner Kirkwood energies for 126 neutrons and 82 protons as 
a function of $\beta_4$. Here,  $\beta_2=0.2$ and $\gamma=0$. Thomas
Fermi energies, second-order corrections and the fourth-order 
corrections are shown in upper, middle and bottom panels respectively.}
\label{figt2}
\end{figure}

A sample WK calculation is performed for system of 126 neutrons and 82 protons. 
The variation of the Thomas-Fermi energy and of the different correction
terms as a function of the quadrupole deformation parameter $\beta_2$ is
presented in Fig. (\ref{figt1}). The other two deformation
parameters, $\beta_4$ and $\gamma$, are set to zero in this test case.
The partial contributions to the WK energy are plotted separately for 
protons and neutrons. It is found that all the correction terms vary 
smoothly as a function of deformation. As expected, the value of
the contributions from the $\hbar^2$ and $\hbar^4$ terms to the averaged energy 
decreases rapidly. It is found that the proton and neutron Thomas-Fermi 
energies have opposite trends with respect to increasing $\beta_2$. 
If Coulomb potential is suppressed, it is found that the Thomas-Fermi
energies for protons follow the same trend as those for the neutrons. 
Further, it is interesting to note that comparatively, the
variation in the second-order corrections with respect to deformation 
parameters is stronger than that in the Thomas-Fermi energies
($\sim 10\%$ for second-order corrections and $\sim 3\%$ for Thomas-Fermi energies).

Next, the variation of the Thomas-Fermi energy and of the correction terms
as a function of the hexadecapole deformation parameter $\beta_4$ is plotted in 
Fig. (\ref{figt2}). Here, $\beta_2$ is taken to be 0.2 and $\gamma$ is
set to zero. It is seen that again, the different energies vary smoothly as a
function of $\beta_4$. The Thomas-Fermi energy for protons is found to have 
very little variation with respect to the $\beta_4$ deformation parameter. 
In contrast, the corresponding energies for neutrons have a stronger dependence 
on $\beta_4$. The same behaviour is also observed in the corresponding quantum 
mechanical energies. It is found that the proton and neutron Thomas-Fermi 
energies have a very similar behaviour if the Coulomb potential is suppressed.
Further, to check if this conclusion depends on the value of 
$\beta_2$, the analysis is repeated for $\beta_2 = 0.4$, and the same 
conclusion is found to emerge.

The behaviour of the Thomas-Fermi energies for protons in the above cases 
(Figs. (2) and (3)) seems to be due to the Coulomb potential. In the 
case of variation with respect to $\beta_2$, qualitatively it can be 
expected that with increasing quadrupole deformation, protons are pulled 
apart and Coulomb repulsion decreases, thereby making the system more bound. 
The $\beta_4$ deformation also affects the proton distribution, but, 
as expected, the effect of hexadecapole deformation is less prominent 
in comparison with that of quadrupole deformation.
Thus, the repulsion among protons does decrease with increasing $\beta_4$, but the 
decrease is not large enough to make the system more bound with larger $\beta_4$. 

By keeping $\beta_2$ and $\beta_4$ fixed, if the parameter $\gamma$ is 
varied, then it is found that the resulting energies are independent of 
the sign of $\gamma$. Moreover, the $\gamma$ dependence of the WK energies
is found to be rather weak. Therefore, here we do not present these result
explicitly.

The fourth-order calculation for protons is very time consuming. Typically, it
takes tens of minutes to do a complete WK calculation. Most of the run-time
being consumed by particle number determination and the fourth-order
calculations for protons. Thus, it is necessary to find an accurate
approximation scheme for the fourth-order calculation for protons. Since in the
nuclear interior, the Coulomb potential has approximately a quadratic nature 
(see Fig. (1)), it
is expected that the Coulomb potential will have small influence on the
fourth-order calculations (note that one needs higher-order derivatives in the
fourth-order energy calculations). One may therefore drop the Coulomb potential completely from the 
fourth-order corrections; we shall refer to this approximation as 
``quadratic approximation''. This approximation has been checked explicitly by
performing exact fourth-order calculations for protons. The maximum
difference between the WK energies obtained by using exact calculation and 
the quadratic approximation is found to be of the order 100 keV for 82 protons.
It turns out that the difference between the quadratic approximation and exact 
calculation decreases with decreasing charge number. This approximation can be improved
by keeping the Laplacian of the Coulomb potential in the
fourth-order contribution i.e., the terms of the form $(\nabla^2 V)^2$
and $\nabla^4 V$ in Eq. (\ref{E_WK}). This means that for protons, 
only the term $\nabla^2\left( \nabla V\right)^2$ is dropped from Eq. (\ref{E_WK}). 
It is found that with this modification, the value of
the fourth-order correction energy for the mean field part for protons almost coincides 
with the value obtained by taking all of the derivatives of the Coulomb potential into account.
This helps in reducing the total runtime further. Thus, effectively, with the interpolation
for Coulomb potential as discussed before (see section IV), and the approximations introduced in 
the fourth-order correction terms for protons in the present section, the runtime 
reduces from tens of minutes to just about two minutes, without affecting the 
desired accuracy of the calculations. 

\section{Wigner-Kirkwood Shell Corrections and Comparison with Strutinsky calculations}

Numerically, it has been demonstrated that the WK and
Strutinsky shell corrections are close to each other \cite{JEN.75}. This is
expected, since it has recently been shown \cite{MED.06} that the Strutinsky
level density is an approximation to the semi-classical WK level density. 
For illustration, we present and discuss the WK and the corresponding 
Strutinsky shell corrections for the chains of Pb, Gd and Dy isotopes.
For the sake of completeness, we first present and discuss the essential
features of the Strutinsky smoothing scheme.

According to the Strutinsky smoothing scheme, the smooth level 
density for a one-body Hamiltonian is given by \cite{BOL.72}:
\begin{eqnarray}
g_{st}(\epsilon)~=~\frac{1}{\gamma\sqrt{\pi}} \sum_{i=1}^{\infty} e^{-(\epsilon - \epsilon_i)^2/\gamma^2} 
                    \sum_{j=1}^{N_s}S_j H_j\left(\frac{\epsilon - \epsilon_i}{\gamma} \right) ,
\end{eqnarray} 
where $\epsilon_i$ are the single-particle energies calculated by 
diagonalising the Hamiltonian matrix. The smoothing constant $\gamma$ is taken 
to be of the order of $\hbar \omega_0$ ($\hbar\omega_0=1.2\times 41 A^{-1/3}$).
$N_s$ is the smoothing order, and is assumed to be equal to 6 in the present
work; $H_j$ are the Hermite polynomials; and $S_j$ is a constant, defined as 
\cite{BOL.72}:
\begin{eqnarray} 
S_j&=&\frac{(-1)^{j/2}}{2^{j}(j/2)!},~~\mathrm{for}~j~\mathrm{even}, \nonumber \\
   &=&0,  \hspace{1.6cm}~~\mathrm{for}~j~\mathrm{odd}.
\end{eqnarray} 
The Strutinsky shell correction is given by:
\begin{eqnarray} 
E_{St}&=&\sum_{i=1}^{N_n}\epsilon_i~-~\int_{-\infty}^{\bar{\lambda}}\epsilon g_{st}(\epsilon)d\epsilon,
\end{eqnarray} 
where $N_n$ is the number of nucleons. This, upon 
substituting the expression for $g_{st}$, yields \cite{BOL.72},
\begin{eqnarray} 
E_{st}&=&\sum_{i=1}^{N_n}\epsilon_i - \sum_{j=0}^{\infty}\left\{\frac{\epsilon_j}{2}\left[1 + \mathrm{erf}(\bar{u}_j)\right] 
 - \frac{\gamma e^{-\bar{u}_{j}^{2}}}{2\sqrt{\pi}} \right. \nonumber \\ 
  && \left. \hspace{1.3cm}- \frac{e^{-\bar{u}_{j}^{2}}}{\sqrt{\pi}}  
\sum_{k=1}^{N_s}S_k \left[\frac{\gamma}{2} H_k(\bar{u}_j) + \epsilon_j H_{k-1}(\bar{u}_j) + 
                          k\gamma H_{k-2}(\bar{u}_j)\right] \right\} ,
\end{eqnarray} 
where
\begin{eqnarray} 
\bar{u}_j~=~\frac{\bar{\lambda}-\epsilon_j}{\gamma} \,.
\end{eqnarray} 
Here, $\bar{\lambda}$ is the chemical potential, calculated iteratively from the
particle number condition. The error integral $\mathrm{erf}(x)$ is defined as:
\begin{eqnarray} 
\mathrm{erf}(x)~=~\frac{2}{\sqrt{\pi}}\int_{0}^{x}e^{-z^2} dz \,.
\end{eqnarray} 

\begin{figure}[htb]
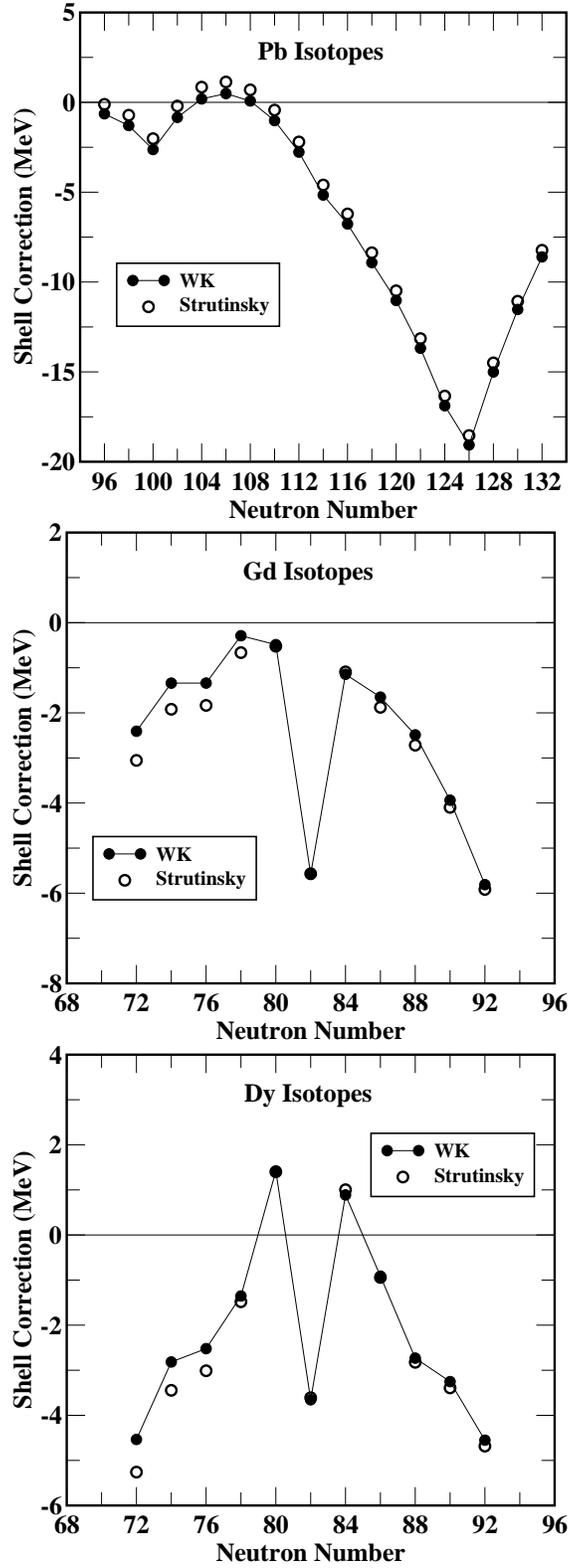

\centerline{\epsfig{file=Pb_compare.eps,width=0.45\textwidth}}
\centerline{\epsfig{file=Gd_compare.eps,width=0.45\textwidth}}
\centerline{\epsfig{file=Dy_compare.eps,width=0.45\textwidth}}
\caption{WK and the corresponding Strutinsky shell corrections
for Pb (upper panel), Gd (middle panel) and Dy (bottom panel) 
isotopes.} 
\label{PbSch}
\end{figure}

It should be noted that the Strutinsky procedure described here uses positive
energy states generated by diagonalizing the Hamiltonian matrix, and not by
taking resonances into account and smoothing them. Further, in practice, the
summations defined above do not extend up to infinity, but are cut off at a
suitable upper limit. The limit is chosen in such a way that all the states up
to $\sim 4 \hbar \omega_0$ are included in the sum. It has been shown that the
uncertainty in the Strutinsky shell corrections obtained this way is typically
of the order of 0.5 MeV \cite{BOL.72}. For lighter nuclei, however, it has been
concluded \cite{BOL.72} that this uncertainty is larger.

The total WK shell correction for the chain of even even Lead isotopes 
($^{178-214}$Pb) is plotted in Fig. (\ref{PbSch}), along with the
corresponding values obtained by using the Strutinsky smoothing method. It is
found that both the WK and Strutinsky results exhibit very similar trends. As
expected, there is a prominent minimum observed for $^{208}$Pb, indicating the
occurrence of shell closure. The WK and Strutinsky shell corrections
slightly differ from each other. The difference is not a
constant, and is found to be increasing slowly towards the more neutron
deficient Lead isotopes. 

Next we plot the calculated (WK) and the corresponding Strutinsky shell
corrections for the chains of even even Gd and Dy isotopes, with neutron numbers ranging 
from 72 to 92. Apart from $^{144,146,148}$Gd and $^{146,148,150}$Dy, the rest of the 
nuclei considered here are known to be deformed \cite{MOL.95}. For this test run, we 
adopt the deformation parameters from the M\"oller - Nix compilation \cite{MOL.95}.
It is seen that the WK and the corresponding Strutinsky shell corrections
agree with each other, within few hundred keVs. The prominent minimum at 
shell closure at neutron number 82 is clearly seen. In these cases as well, 
the difference between the two calculations is not a constant. It is larger 
in the neutron deficient region, and becomes smaller as neutron number increases.

\section{Lipkin-Nogami Pairing Model}
The pairing correlations, important for the open shell nuclei, are often
taken into account within the framework of the BCS model. The BCS model,
however, has two serious shortcomings: 1) particle number fluctuation
(the BCS wavefunctions are not particle number eigenstates), and 2) there
may exist critical values of the pairing strength, below which the
BCS equations may not have any non-trivial solutions. In order to
overcome these difficulties, Lipkin, Nogami
and co-workers proposed to minimise the expectation value of the
model Hamiltonian \cite{LIP.60,NOG.64,HCP.73}:
\begin{eqnarray} 
\hat{\cal{H}} ~=~ \hat{H} - \lambda_1 \hat{N} - \lambda_2 \hat{N}^2
\label{H_LN}
\end{eqnarray} 
by determining $\lambda_1$ and $\lambda_2$ using certain conditions. Here,
$\hat{H}$ is the pairing Hamiltonian, and $\hat{N}$ is the particle number
operator. Minimisation of the expectation value of $\hat{H} - \lambda_1 \hat{N}$
leads to the usual BCS model, with $\lambda_1$ determined from the particle
number condition. Thus, in Eq. (\ref{H_LN}) above, the quantity $\lambda_1$ is a
Lagrange multiplier, but the particle number fluctuation constant
$\lambda_2$ is not.

In practice, the LN calculation is carried out by assuming a constant pairing
matrix element, $G$. For a given nucleus (assumed to be even-even for
simplicity), one considers $N_h$ doubly degenerate states below, and $N_p$
doubly degenerate states above the Fermi level. These states contain $\cal{N}$
nucleons. In practice, one takes $N_h ~=~ N_p ~=~ N/2$ or $Z/2$, depending on
whether it is being applied to neutrons or protons. The occupation probabilities
$v_{k}^{2}$, the pairing gap $\Delta$, the chemical potential $\lambda$ ($=
\lambda_1 + 2\lambda_2({\cal{N}} + 1)$, see Ref.\ \cite{HCP.73}), and the
constant $\lambda_2$ are determined iteratively using the conditions
\cite{HCP.73,MOL.97}:
\begin{eqnarray}
\cal{N}&=& 2\sum_{k} v_{k}^{2} \\
\Delta &=& G\sum_{k} u_k v_k \,,
\end{eqnarray} 
such that
\begin{eqnarray} 
v_{k}^{2}&=&\frac{1}{2} \left[ 1 - \frac{\varepsilon_k - \lambda}
                  {\left\{ \left(\varepsilon_k - \lambda\right)^2 - \Delta^2  \right\}^{1/2}} \right]
\end{eqnarray}
and
\begin{eqnarray}
\varepsilon_k &=& E_k + \left(4\lambda_2 - G\right)v_{k}^{2} \,,
\end{eqnarray} 
where $E_k$ are the single-particle energies and $u_{k}^{2}=1 - v_{k}^{2}$.
The particle number fluctuation constant $\lambda_2$ is given by:
\begin{eqnarray} 
\lambda_2 ~=~ \frac{G}{4} \left[ 
 \frac{ \left(\sum_k u_{k}^{3}v_{k}\right) \left(\sum_k u_{k}v_{k}^{3}\right) - 
\sum_k u_{k}^{4}v_{k}^{4}} 
{\left(\sum_k u_{k}^{2}v_{k}^{2} \right)^{2} - \sum_k u_{k}^{4}v_{k}^{4}}\right]~.
\end{eqnarray} 
The pairing matrix element $G$ is calculated by the M\"oller-Nix 
prescription \cite{MOL.97}:
\begin{eqnarray} 
\frac{2}{G} &=& \bar{\rho}_{L} \ln \left\{ \sqrt{a_{2}^{2} + \bar{\Delta}^{2}} + a_2 \right\} 
              - \bar{\rho}_{L} \ln \left\{ \sqrt{a_{1}^{2} + \bar{\Delta}^{2}} + a_1 \right\}
\end{eqnarray} 
Here, $\bar{\rho}_{L}=g_{WK}/2$ is the Wigner-Kirkwood averaged level density 
(see Eq. (\ref{gWK}). Factor of 2 appears because each quantal level here 
has degeneracy of 2. The level density is evaluated at fermi energy.);
$a_2 = {\cal{N}}/2\bar{\rho}_{L}$ and $a_1 = -{\cal{N}}/2\bar{\rho}_{L}$ and 
$\bar{\Delta}$ is the average pairing gap, taken to be 
$3.3/{\cal{N}}^{1/2}$ \cite{MOL.97}.

The ground-state energy within the LN model is given by:
\begin{eqnarray} 
E_g~=~2 \sum_{k} v_{k}^{2}E_{k} - \frac{\Delta^{2}}{G} - G \sum_{k}v_{k}^{4} 
- 4\lambda_2\sum_{k} u_{k}^{2}v_{k}^{2}~.
\label{E_LN}
\end{eqnarray} 
The pairing correlation energy, $E_{pair}$ is obtained by subtracting the 
ground-state energy in absence of pairing from Eq. (\ref{E_LN}):
\begin{eqnarray} 
E_{pair}~=~E_g - 2 \sum_{k}E_{k} - G{\cal{N}}/2~.
\end{eqnarray} 
 
\section{Calculation of Binding Energies}
As an illustrative example, we now present and discuss the calculated 
binding energies (in this paper, we take binding energies as negative 
quantities) for 367 even-even, even-odd, odd-even and odd-odd spherical 
nuclei. These nuclei are predicted to be spherical or
nearly spherical ($\beta_2 < 0.05$) in the M\"oller-Nix calculations 
\cite{MOL.95} and include $^{38-52}$Ca,
$^{42-54}$Ti, $^{100-134}$Sn, and $^{178-214}$Pb. The detailed list of 
nuclei considered in the present fit can be found in Ref. (\cite{online}). 
Of course, it is
known that the prediction of sphericity does depend to some extent on the
details of the density functional employed \cite{TER.08}. Therefore, it may so
happen that some of the nuclei assumed to be spherical here, may actually turn
out to be slightly deformed when energy minimization is carried out on the grid of deformation parameters.

Our calculation proceeds in the following steps. For each
nucleus, the quantum mechanical and WK energies are calculated as described 
earlier. This then yields values of the 
shell corrections ($\delta E$) for these
nuclei. The pairing energies ($E_{pair}$) are then calculated using the 
Lipkin-Nogami scheme \cite{LIP.60,NOG.64,HCP.73} described previously
in the same potential well where the shell correction is computed. 
These two pieces
constitute the microscopic part of the binding energy. The macroscopic 
part of the binding energy ($E_{LDM}$) is obtained from the liquid drop formula.
Thus, for a given nucleus with $Z$ protons and $N$ neutrons (mass number
$A=N+Z$), the binding energy in the Mic-Mac picture is given by:
\begin{eqnarray} 
E(N,Z)~=~E_{LDM}~+~\delta E ~+~E_{pair}~.
\end{eqnarray} 

The liquid drop part of binding energy is chosen to be:
\begin{eqnarray} 
E_{LDM}&=&a_v\left[1~+~\frac{4k_v}{A^2}~T_z\left(T_z~+~1\right)\right]A 
      ~+~ a_s\left[1~+~\frac{4k_s}{A^2}~T_z\left(T_z~+~1\right)\right]A^{2/3} \nonumber
                     \\
       &+&\frac{3Z^2e^2}{5r_0A^{1/3}}~+~\frac{C_4Z^2}{A} \,,
\end{eqnarray} 
where the terms respectively represent: volume energy, surface energy,
Coulomb energy and correction to Coulomb energy due to surface diffuseness of 
charge distribution. The coefficients $a_v$, $a_s$,
$k_v$, $k_s$, $r_0$ and $C_4$ are free parameters; $T_z$ is the third 
component of isospin, and $e$ is the electronic charge. The free parameters
are determined by minimising the $\chi^2$ value in comparison with the
experimental energies:
\begin{eqnarray}
\chi^2~=~\frac{1}{n}\sum_{j=0}^{n}
      \left[ \frac{ E(N_j,Z_j) - E_{expt}^{(j)} }
                  { \Delta E_{expt}^{(j)}} \right]^2 ,
\end{eqnarray} 
where $E(N_j,Z_j)$ is the calculated total binding energy for the given
nucleus, $E_{expt}^{(j)}$ is the corresponding experimental value 
\cite{WAP.03}, and $\Delta E_{expt}^{(j)}$ is the uncertainty
in $E_{expt}^{(j)}$.
In the present fit, for simplicity, $\Delta E_{expt}^{(j)}$ 
is set to 1 MeV.
%
%
The minimisation is achieved using the well-known
Levenberg-Marquardt algorithm \cite{MAR.68,NR.92}.

\begin{table}[htb]
\caption{Values of the liquid drop parameters obtained through the
$\chi^2$ minimisation.}
\begin{center}
\begin{tabular}{|c|c|c|c|} \hline
{\bf Quantity}&     {\bf Value}          &{\bf Quantity}& {\bf Value}    \\ \hline
 $a_v$        & -15.841 (MeV)            & $a_s$        &~19.173 (MeV)        \\ 
 $k_v$        &  -1.951 \hspace{26pt}    & $k_S$        & -2.577 \hspace{26pt}\\
 $r_0$        & ~~1.187 (fm) \hspace{8pt}& $C_4$        &  1.247 (MeV)        \\ \hline
\end{tabular}
\end{center}
\end{table}

\begin{figure}[htb]
\centerline{\epsfig{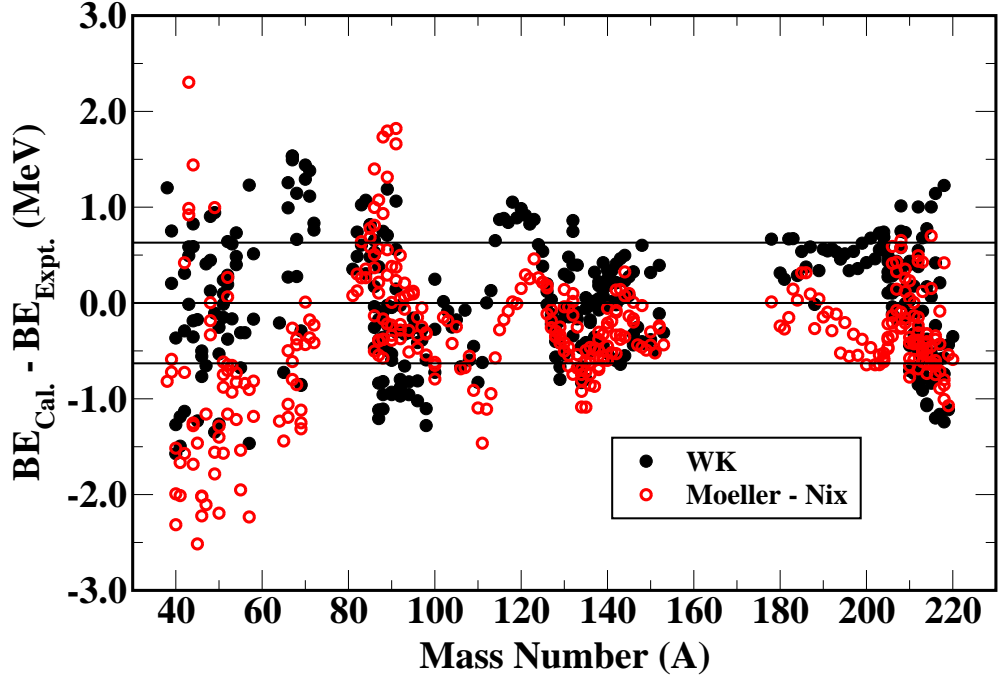}}
\caption{Difference between the calculated and the experimental \cite{WAP.03} 
binding energies. The corresponding differences obtained for the M\"oller-Nix 
values of binding energies \cite{MOL.97} are also presented.
}
\label{diffbe}
\end{figure}

For the set of nuclei considered here, the {\it rms} deviation in binding
energies turns out to be 630 keV, which, indeed is gratifying. The {\it rms}
deviation obtained for the same nuclei with the M\"oller-Nix mass formula 
turns out to be 741 keV. 
The liquid drop parameters are presented in Table~I. Clearly, the obtained
values of the parameters are reasonable. The detailed table containing the
nuclei considered in the present fit, and the corresponding calculated and
experimental  \cite{WAP.03} binding energies may be found in Ref. \cite{online}.

To examine the quality of the fit further, first, we plot the difference between
the fitted and the corresponding experimental \cite{WAP.03} binding energies 
for the 367 nuclei as a function of the mass number A in Fig. 
(\ref{diffbe}) 
The corresponding differences 
obtained for the M\"oller-Nix \cite{MOL.97} values of binding energies are
also plotted in the same figure for comparison. It is amply clear from 
the figure
that the fitted binding energies are close to the experiment (within 1 MeV).
Overall, the quality of the present fit is slightly better than that of the
M\"oller-Nix fit (the {\it rms} deviations, respectively, are 630 and 741 keV). 
Particularly for the lighter nuclei, the present calculations are comparatively 
closer to the experiment.

\begin{figure}[htb]
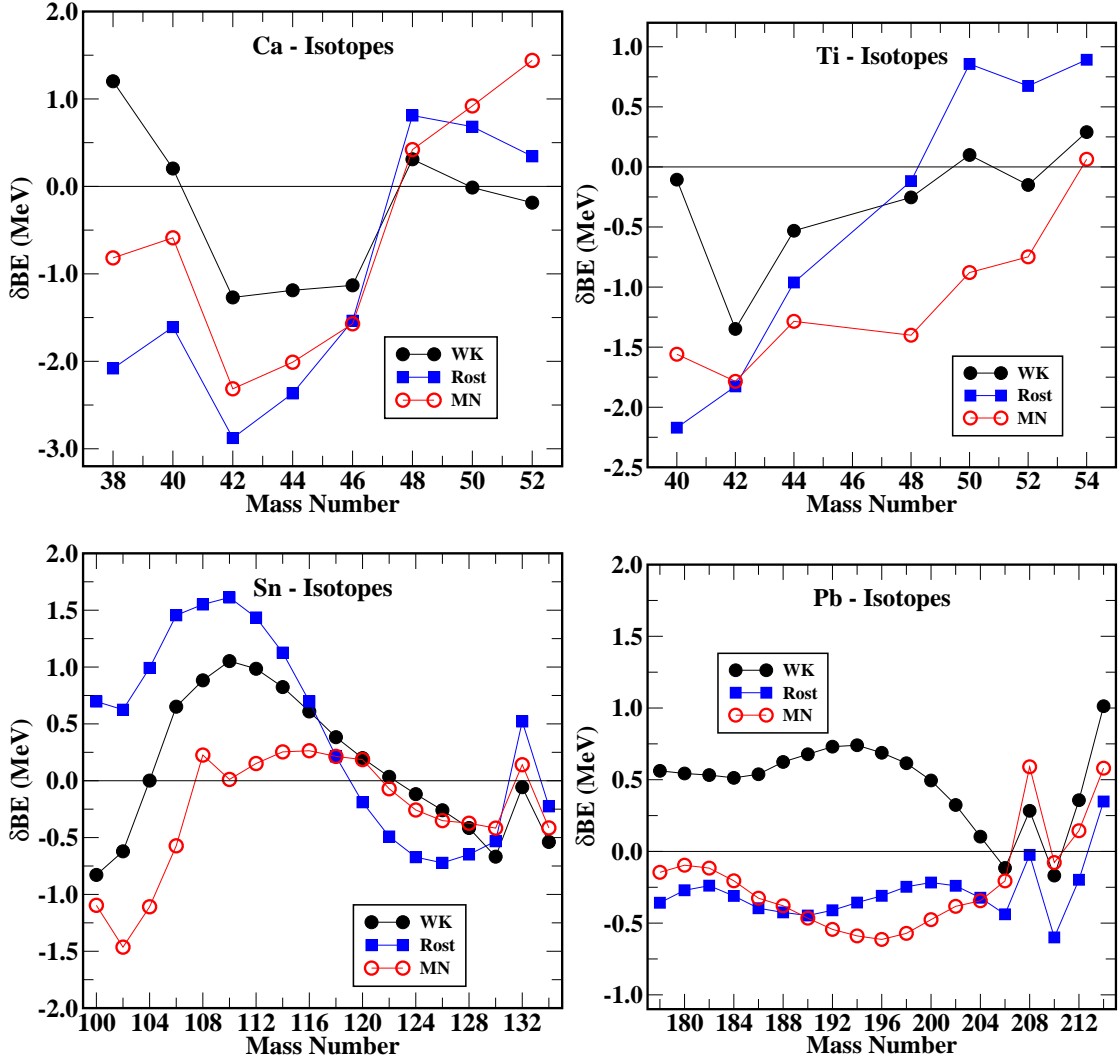

\centerline { \hbox{ \epsfig{file=Ca_be.eps,width=0.45\textwidth} \hfill
                     \epsfig{file=Ti_be.eps,width=0.44\textwidth} }}
\vskip 10pt
\centerline { \hbox{ \epsfig{file=Sn_be.eps,width=0.45\textwidth} \hfill
                     \epsfig{file=Pb_be.eps,width=0.44\textwidth} }}
\caption{
Difference between calculated binding energies and experiment \cite{WAP.03}.
Results are shown for the present calculation (WK), for the M\"oller-Nix
values (MN), and using the Rost parameters in the Woods-Saxon form factors as
described in the text.
}
\label{resu_be}
\end{figure}

Next, the difference between the calculated and the corresponding experimental
\cite{WAP.03} binding energies (denoted by ``WK'') for Ca, Ti, Sn, and Pb
isotopes considered in this fit are presented in Fig. (\ref{resu_be}). The
differences obtained by using the M\"oller-Nix \cite{MOL.97} values of binding
energies (denoted by ``MN'') are also shown there for comparison. It can be seen
that the present calculations agree well with the experiment. It is found that
the differences vary smoothly as a function of mass number: the exceptions being
the doubly closed shell nuclei $^{48}$Ca, $^{132}$Sn, and $^{208}$Pb, where a
kink is observed. The overall behaviour of the differences is somewhat smoother
than that obtained by using the values of M\"oller and Nix.

\begin{figure}[htb]
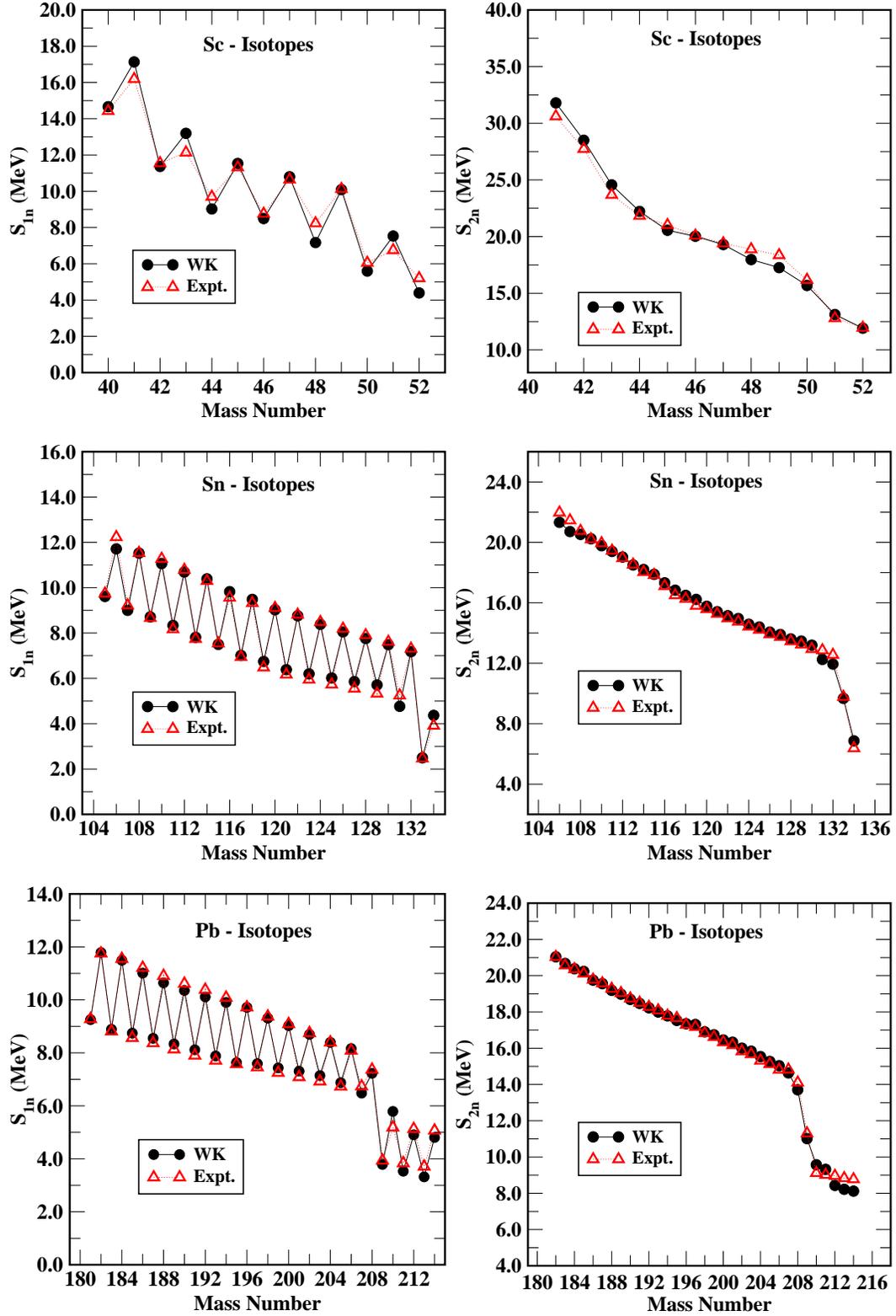

\centerline { \hbox{ \epsfig{file=Sc_s1n.eps,width=0.42\textwidth} \hfill
                     \epsfig{file=Sc_s2n.eps,width=0.42\textwidth} }}
\vskip 10pt
\centerline { \hbox{ \epsfig{file=Sn_s1n.eps,width=0.42\textwidth} \hfill
                     \epsfig{file=Sn_s2n.eps,width=0.42\textwidth} }}
\vskip 10pt
\centerline { \hbox{ \epsfig{file=Pb_s1n.eps,width=0.42\textwidth} \hfill
                     \epsfig{file=Pb_s2n.eps,width=0.42\textwidth} }}
\caption{The calculated and the corresponding experimental \cite{WAP.03}
one and two-neutron separation energies for Sc, Sn and Pb isotopes.}
\label{pb_s2n}
\end{figure}

To investigate the effect of the parameters of the single-particle potential,
we make a refit of the liquid drop parameters, by using the Rost parameters
\cite{ROS.68} in the microscopic part of the binding energy computed in 
the WK approximation. That is, we
calculate the shell corrections and pairing energies employing the Rost
parameters for the Woods-Saxon form factors, and then fit the liquid drop
parameters again for the same set of nuclei. The {\it rms} deviation obtained
in this case (1.14 MeV for even even nuclei) is much worse that the one 
obtained for the parameters
mentioned in Section 3 (the {\it rms} deviation obtained for this potential 
is around 0.680 MeV), which is amply clear from the figure. It is well known
that the Rost parameters have very large half density radii. As a consequence,
the values of the moment of inertia and {\it rms} radii obtained by using
these parameters deviate strongly from the corresponding experimental values.
On the contrary, the parametrisation used in the present analysis (see Section
3), yields reasonable values of moments of inertia and radii. Thus, overall,
this potential is more realistic than the Rost potential. This is reflected in
the calculated binding energies as well, showing clearly that the choice of
the single-particle potential (or in other words, the parameters) is indeed
important for reliable predictions of binding energies (and hence the masses).

Single and two neutron separation energies ($S_{1n}$ and $S_{2n}$) 
are crucial observables. They are obtained by calculating binding energy differences 
between pairs of isotopes differing by one and two neutron numbers, respectively. 
The single neutron separation energies govern asymptotic behaviour
of the neutron density distributions \cite{YKG.85}. They exhibit odd-even 
staggering along an isotopic chain, indicating that the isotopes with even
number of neutrons are more bound than the neighbouring isotopes with odd
number of neutrons. The systematics of $S_{2n}$ primarily reveals the shell 
structure in an isotopic chain. The correct prediction of these separation 
energies is crucial for determination of the neutron drip lines.
The calculated $S_{1n}$ and $S_{2n}$ values for Sc, Sn and Pb isotopes are displayed 
in Fig. (\ref{pb_s2n}). The corresponding experimental values
of $S_{1n}$ and $S_{2n}$ \cite{WAP.03} are also plotted for comparison. The agreement 
between calculations and experiment is found to be excellent. The odd - even staggering 
is nicely reproduced. The shell closures at $^{132}$Sn and $^{208}$Pb are clearly
visible both in single and two neutron separation energies. At a finer
level, however, a marginal underestimation of the shell gap at the neutron
number 82 (126) is observed in $^{132}$Sn ($^{208}$Pb). Finally, we remark that the
calculated single and two proton separation energies are also found to be in 
close agreement with the experiment. 

The results presented in this section indicate that the present calculations of 
binding energies, indeed, are reliable.

\section{Summary and future outlook}
In the present work, we intend to carry out reliable mass 
calculations for the nuclei spanning the entire periodic table.
For this purpose, we employ the `microscopic-macroscopic'
framework. The microscopic component has two ingredients:
the shell correction energy and the pairing energy. The pairing energy is
calculated by using the well-known Lipkin-Nogami scheme.
To average out the given one-body Hamiltonian (and hence
find the shell corrections, given the total quantum mechanical
energy of the system), we use the semi-classical Wigner-Kirkwood 
expansion technique. This method does not use 
the detailed single-particle structure, as in the case of 
the conventional Strutinsky smoothing method. In addition to
the bound states, the Strutinsky scheme requires the contributions
coming in from the continuum as well. Treating the continuum is often tricky,
and in most of the practical calculations, the continuum is 
taken into account rather artificially, by generating 
positive energy states by means of diagonalisation of the 
Hamiltonian matrix. For 
neutron-rich and neutron-deficient nuclei, the contribution from
the continuum becomes more and more important as the 
Fermi energy becomes smaller (less negative). Uncertainty
in the conventional Strutinsky scheme thus increases as 
one goes away from the line of stability. It is 
therefore expected that the Wigner-Kirkwood method will be 
a valuable and suitable option especially for nuclei lying far away from the
line of stability.

We now summarise our observations and future perspectives:
\begin{enumerate}
\item Semi-classical averaging of a realistic one-body Hamiltonian
using the Wigner-Kirkwood expansion of the partition function
correct up to fourth order in $\hbar$ is carried out for the 
deformed systems, both for protons and neutrons. 
The spin-orbit as well as Coulomb potentials are explicitly 
taken into account.
\item The smooth energies thus obtained are investigated in 
detail as a function of three deformation parameters:
$\beta_2$, $\beta_4$, and $\gamma$. As expected, the energies
corresponding to the leading-order term in the expansion as well
as the correction terms vary smoothly as a function of
deformation parameters.
\item Differences between the quantum mechanical and the corresponding 
averaged energies yield the shell corrections. These, along with the pairing
energies obtained by using the Lipkin-Nogami scheme constitute the
``microscopic'' part of the nuclear binding energy in the `Mic-Mac' 
picture. 
Using a simple liquid drop ansatz with six adjustable parameters,
it is demonstrated that the present approach indeed, is feasible, 
and very promising. For the test case presented here, comprising 
of 367 spherical nuclei, the {\it rms} deviation of the predicted 
binding energies from the experimental values turns out to be 
630 keV.
\item The importance of the one-body potential in reliable estimations of
nuclear binding energies is explicitly demonstrated. It should be 
noted that the Woods-Saxon parameters used in this work have been
fitted for the Coulomb potential calculated by using the uniform density
(sharp surface) approximation. The Coulomb potential we use is 
obtained from folding Woods-Saxon density profile with the Coulomb
interaction. Therefore, before performing the large scale calculations, 
we intend to make a refit to the Woods-Saxon potential, with the Coulomb 
potential obtained from folding.
\item Having established the feasibility of the present approach,
we now intend to extend our binding energy calculations to deformed
nuclei. For this purpose, we plan to minimise the binding energy on a
mesh of deformation parameters to find the absolute minimum in 
the deformation space. Work along these lines is in progress.
 
\end{enumerate}

\appendix

\section{Geometry of Distance Function}
\begin{figure}[htb]
\centerline{\epsfig{file=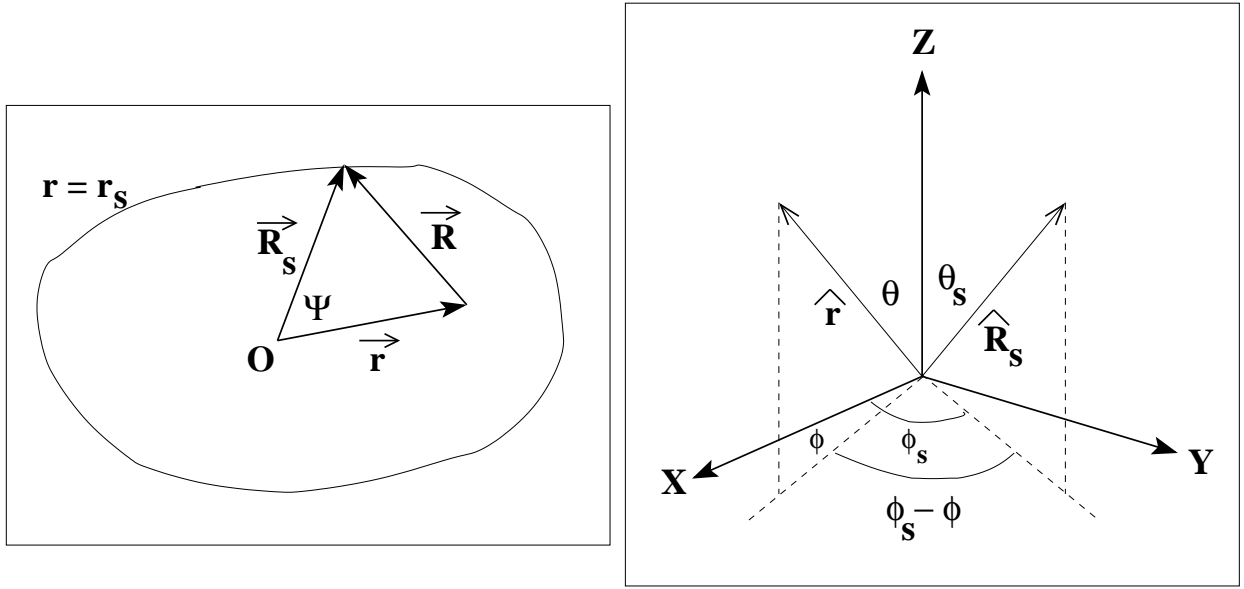,width=\textwidth}}
\label{geom}
\caption{Geometry of distance function}
\end{figure}
Consider an arbitrary surface, defined by the relation
$r~=~r_s$, where $r_s$ is given by Eq. (\ref{def}) of the text. Let us fix
the origin of the coordinate system at the centre of mass of the object.
Let $\vec{r}\equiv (r,\theta,\phi)$ define an arbitrary point in space. 
This point could be inside or outside the surface. Here, for concreteness, 
we assume that it is within the volume of the object. Our aim is 
to find the minimum distance of the point $\vec{r}$ to the surface $r~=~r_s$.
To achieve this, construct a vector $\vec{R}_s$
from the centre of mass to the surface. To find the minimum distance,
one has to minimise the object $|\vec{r}-\vec{R}_s|$. Denoting the 
angle between the $\vec{r}$ and $\vec{R}_s$ vectors as $\Psi$, we have:
\begin{eqnarray} 
|\vec{r}-\vec{R}_s|~=~\sqrt{R_{s}^{2} + r^2 - 2rR_s\cos\Psi}~,
\end{eqnarray} 
where, from Fig. (\ref{geom}), the cosine of the angle $\Psi$ is
given by:
\begin{eqnarray} 
\cos\Psi~=~\cos\theta\cos\theta_s+\sin\theta\sin\theta_s\cos(\phi_s-\phi) .
\end{eqnarray}
The latter result can be proved easily by considering a unit sphere, and a
spherical triangle constructed with unit vectors $\hat{r}$, $\hat{R}_s$, and
$\hat{z}$. For spherical symmetry, the vectors $\vec{r}$ and $\vec{R}_s$ are
parallel to each other, and one recovers the usual spherical Woods-Saxon form
factor. 

\section{Coulomb Potential and its Derivatives}
\subsection{Proof of Eq. (\ref{altcoul})}
The Coulomb potential for an arbitrary charge distribution is
given by:
\begin{eqnarray} 
V_C(\vec{r})~=~e^2\int \rho(\vec{r}') \frac {1} {|\vec{r}-\vec{r}'|} d\vec{r}'~.
\end{eqnarray}
Let, for brevity, $|\vec{r}-\vec{r}'|~=~{\cal {R}}$. Consider:
$$ \vec{\nabla}_{\vec{r}'}\left\{ \frac{\vec{r}-\vec{r}'} {{\cal {R}}} \right\}$$
Here, the symbol $\vec{\nabla}_{\vec{r}'}$ means that the differentiation is 
done with respect to the $r',\theta',\phi'$ coordinates. Let us consider the above
derivative component-wise. The contribution coming from the first component 
is:
\begin{eqnarray} 
\partial_{x^{'}_{1}} \frac {x_1 - x^{'}_{1} } { {\cal {R}} } ~=~ \frac{-1}{{\cal {R}}} + 
\frac{ \left( x_1 - x^{'}_{1} \right)^2} { {\cal {R}}^{3/2} }
\end{eqnarray} 
Adding contributions coming from all the three components, one gets:
\begin{eqnarray} 
\vec{\nabla}_{\vec{r}'} \cdot \left\{ \frac{\vec{r}-\vec{r}'} {{\cal {R}}} \right\} ~=~ \frac{-2}{{\cal {R}}}
\end{eqnarray} 
With this, the potential becomes:
\begin{eqnarray} 
V_C(\vec{r})&=&\frac{-e^2}{2}\int \rho(\vec{r}') \vec{\nabla}_{\vec{r}'} \cdot \left\{ \frac{\vec{r}-\vec{r}'} {{\cal {R}}} \right\} d\vec{r}' \\
            &=&\frac{-e^2}{2}\int \rho(\vec{r}') \vec{\nabla}_{\vec{r}'} \cdot   \hat{ {\cal {R}} } d\vec{r}'
\end{eqnarray} 
Here, the term in the curly brackets has been represented as a unit vector 
$\hat{ {\cal {R}} }$. Using the identity:
\begin{eqnarray} 
\vec{\nabla}_{\vec{r}'} {\cal {R}} ~=~ -\hat{ {\cal {R}} }~,
\end{eqnarray} 
one obtains:
\begin{eqnarray} 
V_C(\vec{r})~=~\frac{e^2}{2} \int \rho(\vec{r}') \nabla^{2}_{\vec{r}'}  |\vec{r}-\vec{r}'| d\vec{r}'
\end{eqnarray} 
which, upon integrating by parts and transferring derivatives to density, 
becomes:
\begin{eqnarray} 
V_C(\vec{r})~=~\frac{e^2}{2}\int d\vec{r}' |\vec{r}-\vec{r}'| \nabla^{2}_{\vec{r}'} \rho(\vec{r}')
\end{eqnarray} 
\begin{flushright} q.e.d. \end{flushright}

\subsection{Derivatives of Coulomb Potential}
The calculation of the higher-order derivatives of the Coulomb potential
(third and above), even with the form defined in Eq. (\ref{altcoul}), turns
out to be numerically unstable. For this purpose, we employ the Poisson's 
equation. According to this, the Laplacian of Coulomb potential is
proportional to the charge density:
\begin{eqnarray} 
\nabla^2 V_C(\vec{r})~=~-4\pi e^2 \rho(\vec{r})
\end{eqnarray} 
The Laplacian of $\nabla^2 V_C(\vec{r})$ is simple to compute, 
for, all one needs to calculate there are the derivatives of 
density (assumed to be of Woods-Saxon form). 

Thus, it is desirable to generate the required higher order-derivatives
of the Coulomb potential (see expression (\ref{cross}) in the text)
from Poisson's equation. For this purpose, 
we evaluate the commutators:
\begin{eqnarray} 
\left[\nabla^2 \frac{\partial}{\partial r},\frac{\partial}{\partial r}\nabla^2\right]V_C(\vec{r}),
\left[\nabla^2 \frac{1}{r}\frac{\partial}{\partial \theta},\frac{1}{r}\frac{\partial}{\partial \theta}\nabla^2\right]V_C(\vec{r}),
\left[\nabla^2 \frac{\csc \theta}{r}\frac{\partial}{\partial \theta},\frac{\csc \theta}{r}\frac{\partial}{\partial \theta}\nabla^2\right]V_C(\vec{r})
\nonumber
\end{eqnarray} 
The results are:
\begin{eqnarray} 
\nabla^2 \frac{\partial}{\partial r}V_C & = & \frac{\partial}{\partial r}\nabla^2 V_C 
+ \frac{2}{r} \left[ \nabla^2 V_C - \frac{1}{r}\frac{\partial}{\partial r} V_C -\frac{\partial^2}{\partial r^2}V_C \right] \\
\nabla^2 \left( \frac{1}{r} \frac{\partial}{\partial \theta}V_C\right) & = & 
 \frac{1}{r} \frac{\partial}{\partial \theta}\nabla^2 V_C 
+ \frac{\csc^2\theta}{r^3}\left[  \frac{\partial}{\partial \theta}V_C + 2\cot\theta \frac{\partial^2}{\partial \phi^2}V_C \right]
 -  \frac{2}{r^2} \frac{\partial^2}{\partial \theta \partial r} V_C \\
\nabla^2 \left( \frac{\csc\theta}{r} \frac{\partial}{\partial \phi}V_C\right) & = &
\frac{\csc\theta}{r} \frac{\partial}{\partial \phi}\nabla^2V_C + \frac{\csc^3\theta}{r^3} \frac{\partial}{\partial \phi}V_C
- \frac{2\csc\theta}{r^2} \left[ \frac{\cot\theta}{r} \frac{\partial^2}{\partial \theta \partial\phi} V_C 
+ \frac{\partial^2}{\partial \theta \partial r} V_C \right] \nonumber \\
\end{eqnarray} 
With these expressions, the required higher-order derivatives of the Coulomb
potential can be generated. These are then used to evaluate the fourth-order
WK energy, as we have described in Section 4.

\begin{acknowledgments}
AB is thankful to KTH Stockholm and IPN Orsay for financial support.
M.C. and X.V. partially supported by the Consolider Ingenio
2010 Programme CPAN CSD2007-00042 and grants FIS2008-01661 from MEC
and FEDER and 2009SGR-1289 from Generalitat de Catalunya. 
\end{acknowledgments}

\end{document}